\documentclass[a4paper,12pt]{article}

\usepackage[english]{babel}
\usepackage{fullpage}

\usepackage{amsmath, amssymb, amsthm}
\usepackage{verbatim}

\usepackage{xcolor}

\usepackage{graphicx}

\allowdisplaybreaks[3]

\usepackage[hang]{footmisc}
\usepackage[compress,square,numbers]{natbib}
\usepackage[colorlinks,linktocpage,linkcolor=blue,citecolor=purple,urlcolor=purple]{hyperref}

\def\del {\partial}
\def\d {{\rm d}}

\newcommand{\R}{\mathcal{R}}
\newcommand{\nn}{\nonumber}

\newcommand{\e}{\mathrm{e}}
\newcommand{\w}{\wedge}

\newcommand{\nl}{\notag \\ &\quad\,}
\newcommand{\nll}{\notag \\ &}

\definecolor{darkgreen}{rgb}{0.0, 0.42, 0.24}

\begin{document}
\numberwithin{equation}{section}
\thispagestyle{empty}

\vspace*{1cm}

\begin{center}

{\LARGE \bf{Almost classical de Sitter?}}\\

\vspace{0.9 cm} {\large Ludwig Horer$^{1,2}$, Daniel Junghans$^{1}$}\\
\vspace{0.9 cm} {\small\slshape $^1$ Institute for Theoretical Physics, TU Wien\\
Wiedner Hauptstrasse 8-10/136, A-1040 Vienna, Austria}\\
\vspace{0.2 cm} {\small\slshape $^2$ Laboratoire d’Annecy-le-Vieux de Physique Th\'eorique (LAPTh),\\
CNRS, Universit\'e Savoie Mont Blanc (USMB),\\
9 Chemin de Bellevue, 74940 Annecy, France}\\
\vspace{0.9cm} {\upshape\ttfamily ludwig.horer@tuwien.ac.at, daniel.junghans@tuwien.ac.at}\\
\vspace{0.9cm} \today

\vspace{0.5cm}

\begin{abstract}
\noindent The classical-dS scenario in the type II string theories proposes to search for dS vacua of orientifold flux compactifications in a regime where string corrections to the compactified effective field theory are negligible. We study a minimal extension of this scenario in which the leading string corrections to the O-plane/D-brane actions at the 4-derivative order are included but higher orders as well as string corrections in the bulk are self-consistently neglected.
Our proposal is motivated by a recent debate about dS solutions with O8-planes which circumvent a classical no-go theorem due to unusual sources leading to so-called permissive boundary conditions for the 10D supergravity fields.
We argue that such sources do not arise in classical supergravity but ask whether including the 4-derivative corrections leads to sources that have a similar effect. However, we find that the 4-derivative corrections do not allow meta-stable dS in a class of models with O8-planes and/or D8-branes we consider.
We also study related models which in addition contain O6-planes/D6-branes and find that again no meta-stable dS is allowed, both classically and including the 4-derivative corrections. While some of the arguments in this work require the backreaction of the O-plane/D-brane sources to be small, others are valid including the full backreaction.

\end{abstract}

\end{center}

\newpage

\tableofcontents

\section{Introduction}

If string theory has anything to do with the real world, it must be able to explain the accelerated expansion of our universe. The simplest way to achieve this would be a compactification to dS space. However, in spite of two decades of intense research, it is still unclear whether string theory admits such compactifications.
Although there are a number of popular scenarios such as \cite{Kachru:2003aw,Balasubramanian:2005zx}, no fully explicit dS models have been constructed so far.
Furthermore, various arguments and conjectures have been put forward which express doubts about the consistency of dS space in quantum gravity \cite{Danielsson:2018ztv, Obied:2018sgi, Ooguri:2018wrx, Dvali:2017eba, Dvali:2018jhn}, and there is growing evidence that realizing dS space in string theory may not be possible in regimes where string corrections are self-consistently controlled \cite{Junghans:2018gdb, Banlaki:2018ayh, Carta:2019rhx, Gao:2020xqh, Blumenhagen:2022dbo, Junghans:2022exo, Gao:2022fdi, Junghans:2022kxg, Hebecker:2022zme, Schreyer:2022len, Schreyer:2024pml, ValeixoBento:2023nbv, Junghans:2023lpo}.

A fairly explicit setting in which one can look for dS vacua is the classical-dS scenario pioneered in \cite{Hertzberg:2007wc, Silverstein:2007ac}, where one considers orientifold compactifications with background fluxes in the classical regime of type IIA/B string theory (see, e.g., \cite{Danielsson:2011au, Andriot:2020wpp, Andriot:2022way, Andriot:2022yyj} for the state of the art and further references).
Here by ``classical regime'' we mean that the lower-dimensional effective field theory in which the putative dS vacua arise is obtained by dimensionally reducing the 10D type IIA/B supergravity action at the two-derivative level, supplemented by the leading-in-$\alpha^\prime$ actions of localized sources (branes and O-planes). A necessary consistency condition of such a classical approximation is that the ignored perturbative and non-perturbative string corrections must be negligibly small
in the lower-dimensional theory (but not necessarily at every point in the 10D spacetime). One can argue that this is the case at a sufficiently large volume and small string coupling.

The classical-dS scenario is attractive since it potentially yields very explicit models which in contrast to other scenarios neither require multiple steps of moduli stabilization and uplifting at different energy scales nor rely on ingredients like instanton corrections
whose full moduli dependence is hard to compute. The downside is that no-go theorems rule out dS extrema in many classical models, see, e.g., \cite{Hertzberg:2007wc, Wrase:2010ew}. Moreover, the few dS extrema that have been found in the literature are perturbatively unstable and have $\mathcal{O}(1)$ curvature and string coupling when flux/charge quantization and tadpole constraints are correctly imposed.
The ubiquitous instabilities
are often explained in terms of a universal tachyon \cite{Danielsson:2012et, Junghans:2016uvg, Junghans:2016abx}. Furthermore, scaling arguments suggest that getting classical dS at small curvature and small coupling is in general impossible in broad classes of models \cite{Junghans:2018gdb, Banlaki:2018ayh}.\footnote{Possible counter-examples of (unstable) dS solutions admitting scaling limits with parametrically small curvature were very recently proposed in \cite{Andriot:2024cct}. However, these solutions may receive large backreaction corrections from O5-planes which effectively have codimension 2 in these limits.} However, there is so far no rigorous proof ruling out the scenario completely (see, e.g., \cite{Andriot:2019wrs} for a discussion).

A few years ago, the classical-dS idea received renewed attention due to the work \cite{Cordova:2018dbb} which
claimed to have found classical dS vacua in a strikingly simple model (which we will call the \emph{CDT1 model} in this paper).
Specifically, the authors considered a compactification of type IIA string theory on a negatively curved Einstein space times a circle with two parallel O8-planes as the only localized sources and Romans mass as the only flux. The equations of motion in this model reduce to a few simple ODEs which can be solved explicitly including the full non-linear backreaction of the O-planes. This simplicity was surprising since all previously studied models which had not been ruled out by no-go theorems were much more complicated and involved, e.g., intersecting O-planes, NSNS and/or RR fluxes of several ranks as well as geometric fluxes.

While the CDT1 model
seemed promising, it was subsequently shown in \cite{Cribiori:2019clo} that its simplicity comes at a price. Indeed, the localized sources in the model appear in the 10D supergravity equations in a way which is not expected from the standard O8-plane action at leading order in $\alpha^\prime$ so that it is unclear whether the sources have a string-theory interpretation as genuine O8-planes. More generally, \cite{Cribiori:2019clo} showed that the equations of motion derived from the classical type IIA supergravity action and the leading-in-$\alpha^\prime$ O8-plane/D8-brane action rule out classical dS vacua in \emph{every} flux compactification (i.e., for arbitrary compact space and fluxes) in which only O8-planes and D8-branes but no further localized sources with higher codimension are present.
A similar no-go theorem was also proven in the earlier work \cite{Andriot:2016xvq} assuming a specific ansatz for the dilaton. See also \cite{Kim:2020ysx, Bena:2020qpa} for related works studying aspects of the CDT1 model.

A possible loophole to the dS no-go of \cite{Cribiori:2019clo} was pointed out in \cite{Cordova:2019cvf} based on the observation of apparent ambiguities in the supergravity equations (so-called \emph{permissive} boundary conditions) that would allow source terms violating the assumptions in \cite{Cribiori:2019clo}. We will argue below and in upcoming work \cite{Junghans:2024} that these ambiguities do not arise at the level of the classical supergravity approximation and that instead one should obtain precisely the sources assumed in the no-go. However,
as pointed out in \cite{Cribiori:2019clo, Cordova:2019cvf}, $\alpha^\prime$ corrections may change this conclusion. In particular, \cite{Cribiori:2019clo} proposed to circumvent the no-go by turning on $\alpha^\prime$ corrections to the O8/D8 action. Such corrections are known to arise at the 4-derivative order (see, e.g., \cite{Bachas:1999um, Wyllard:2000qe, Wyllard:2001ye, Fotopoulos:2001pt, Schnitzer:2002rt, Garousi:2006zh, Garousi:2009dj, Robbins:2014ara, Garousi:2014oya}). Taking into account these terms provides further couplings of the O8/D8 to the bulk fields and thus modifies the way the latter are sourced in the equations of motion.
An intriguing possibility is therefore that adding a finite number of leading higher-derivative couplings could suffice to get dS vacua in the CDT1 model without requiring complicated new ingredients in the bulk or sacrificing perturbative control.

The purpose of the present work is to study this idea in detail.
In particular, we propose a minimal extension of the classical-dS scenario where we include the effect of the leading $\alpha^\prime$ corrections on the O-plane/D-brane actions at the 4-derivative level but still require that corrections with more than 4 derivatives are negligible.
One may think that this approach cannot make sense since O-planes are often surrounded by ``holes'' in which the $\alpha^\prime$ expansion breaks down. However, in the regime where these holes are small and supergravity is valid on most of the spacetime, we will argue that the O-plane contribution to the vacuum energy is indeed dominated by the classical source terms and their 4-derivative corrections whereas the effect of the non-perturbative hole region only plays a role for the short-distance physics.

We call this scenario the \emph{almost-classical-dS scenario} as it only mildly modifies the classical-dS one while retaining most of its nice properties. In particular, the 10D bulk equations of motion are still classical and given in the case at hand by the simple ODEs derived in \cite{Cordova:2018dbb}. The only change compared to the classical scenario is that the boundary conditions at the O8/D8 loci are now modified due to the new couplings from the $\alpha^\prime$ corrections we turn on. A reasonable hope is that this is sufficient to get dS solutions similar to those in \cite{Cordova:2018dbb} and thus resolve the issues reported in \cite{Cribiori:2019clo}.
However, one of the main results of this paper is that this does in fact not work: Indeed, we will argue that the $\alpha^\prime$ corrections to the O8/D8 actions cannot yield meta-stable dS solutions.
Relaxing the requirement of classical O-plane/D-brane actions in the controlled way we propose does therefore not help to circumvent the no-go theorem of \cite{Cribiori:2019clo}.

Another goal of this work is to analyze a second classical-dS model (called \emph{CDT2 model} from now on), which was proposed in \cite{Cordova:2019cvf} and involves both O6-planes and O8-planes. Although the CDT2 model is slightly more complicated than the CDT1 model,
the equations of motion are still very simple and in particular again reduce to a set of ODEs which can be solved explicitly including the full non-linear O-plane backreaction. As already noted in \cite[footnote 8]{Cordova:2019cvf}, the dS solutions in this model have sources which differ in some aspects from the expected behavior of O6-planes. A natural question is therefore whether
the CDT2 model suffers from similar issues as reported in \cite{Cribiori:2019clo} for the CDT1 model.

We will find in this work that this is indeed the case.
In particular, we will show that the equations of motion obtained from the classical type IIA bulk action and the leading-in-$\alpha^\prime$ O8/O6-plane actions only admit AdS but no dS solutions. Similar conclusions apply to certain variants of the CDT2 model in which we relax some of the assumptions of \cite{Cordova:2019cvf} on the geometry, the fluxes and the sources. We will argue that, if dS solutions exist in such ``CDT2-like'' models, they are unstable. We will further show that these problems cannot be remedied by including 4-derivative couplings, as their effect on the classically non-vanishing vacuum energy and moduli masses can be shown to be subleading in the controlled regime where the curvature is small on most of the spacetime.

Let us note here that some of our arguments in this paper -- in particular those identifying an instability -- assume the smeared approximation. We will explain and justify this assumption in more detail below. We also stress that our no-go arguments for the CDT2 model are valid including the full backreaction. More results taking into account the backreaction in the CDT1 model will be provided in \cite{Junghans:2024}. See Table \ref{tab1} for a summary of the assumptions we impose for our different no-gos.

This paper is organized as follows. In Section \ref{sec:setup}, we set the stage by stating our conventions and some basic facts about type IIA flux compactifications with localized sources. We also discuss the different assumptions on O8 sources and boundary conditions in this paper and the previous works \cite{Cordova:2018dbb, Cribiori:2019clo, Cordova:2019cvf}, and we review the logic behind the smeared approximation, which will be useful for our analysis in this work.
In Section \ref{sec:o8}, we review the classical-dS no-go of \cite{Cribiori:2019clo} for the CDT1 model and other models with O8/D8 sources and argue that it cannot be circumvented by including 4-derivative $\alpha^\prime$ corrections.
Along the way, we clarify some subtleties in the original argument of \cite{Cribiori:2019clo} and address related criticisms in \cite{Cordova:2019cvf}, which we believe are not justified.
In Section \ref{sec:o6}, we derive a similar classical-dS no-go in the CDT2 model and variants thereof and show that the 4-derivative $\alpha^\prime$ corrections are again not sufficient to circumvent this no-go. We conclude with a discussion of our results in Section \ref{sec:concl}. The classical type IIA equations of motion are stated in App.~\ref{app:eom0}, and a computation of O6-plane boundary conditions relevant for Section \ref{sec:o6} is detailed in App.~\ref{app:o6}.

\section{Preliminaries}
\label{sec:setup}

In this section, we review some basic facts and recent results about type IIA flux compactifications with localized sources that will be relevant for this paper.
We first state the action and our general ansatz in Section \ref{sec:setup1}.
We then discuss in Section \ref{sec:bc} the different assumptions on O8 sources and boundary conditions used in this paper and the previous works \cite{Cordova:2018dbb, Cribiori:2019clo, Cordova:2019cvf}. In Section \ref{sec:smear}, we provide a detailed discussion of backreaction effects and review the logic behind the smeared approximation.

\subsection{Type IIA supergravity}
\label{sec:setup1}

The bosonic part of the type IIA supergravity action in the string frame and the democratic formulation \cite{Bergshoeff:2001pv} is
\begin{equation}
S = \frac{1}{2 \kappa^2} \int \d^{10} x \sqrt{-g} \left( \e^{-2\phi} \left(\mathcal{R} + 4 (\del \phi)^2 - \frac{1}{2} |H_3|^2 \right) - \frac{1}{4} \sum_{q=0,2,\ldots}^{10} |F_q|^2 \right) + S_{\text{sources}} \label{action}
\end{equation}
with $2\kappa^2 = (2\pi)^7\alpha^{\prime 4}$. The NSNS and RR field strengths are given by $H_3=\d B_2$ and $F_q = \d C_{q-1}-H_3\w C_{q-3}+F_0\,\e^{B_2}|_q$ in terms of the gauge potentials $B_2$ and $C_{q-1}$ in local patches away from the localized sources. The RR field strengths further satisfy the duality relation $\star_{10} F_q = (-1)^{q(q-1)/2} F_{10-q}$ which needs to be imposed on-shell.

The sources we consider are O$p_i^\pm$-planes and D$p_i$-branes wrapped on $(p_i+1)$-dimensional submanifolds $\Sigma_i$ with the action\footnote{We ignore couplings to $B_2$ and the worldvolume gauge field in the leading D-brane action, which are not relevant for this work.
}
\begin{equation}
S_\text{sources} =  - \sum_i T_{i} \left(\int_{\Sigma_i} \d^{p_i+1} x \sqrt{-g_{p_i+1}}\, \e^{-\phi} \left(1 + \mathcal{L}_{\alpha^{\prime 2},i}\right) - \int_{\Sigma_i} C_{p_i+1}\right), \label{action2}
\end{equation}
where $T_{i}\in\{T_\text{O$p_i^\pm$},T_\text{D$p_i$}\}$ is the tension/charge satisfying
\begin{equation}
T_\text{O$p_i^\pm$} = \pm 2^{p_i-4}T_\text{D$p_i$}, \qquad T_\text{D$p_i$} = \frac{2\pi}{(2\pi\sqrt{\alpha^\prime})^{p_i+1}} \label{tension}
\end{equation}
on the covering space of the orientifold.
The corresponding action of an anti-O$p_i^\pm$-plane or anti-D$p_i$-brane is obtained by flipping the sign of the last term in \eqref{action2}.
The models studied in this paper have sources with $p_i=6,8$.

In \eqref{action2}, we included the next-to-leading-order corrections to the O-plane/D-brane actions at the 4-derivative level in the $\alpha^\prime$ expansion, which are schematically of the form
\begin{align}
\mathcal{L}_{\alpha^{\prime 2},i} = (2\pi)^4\alpha^{\prime 2}& \Big( c_{1i} \e^{4\phi} F_0^4 + c_{2i} \e^{2\phi}F_0^2 \mathcal{R} + c_{3i} \mathcal{R}^2 + c_{4i} H_3^4 + c_{5i} H_3^2 \mathcal{R} + c_{6i} \e^{2\phi} F_0^2 H_3^2 \nll
+ c_{7i} \e^{4\phi} F_2^4 + c_{8i} \e^{4\phi} F_0^2 F_2^2 + c_{9i} \e^{2\phi} F_2^2 \mathcal{R} + c_{10i} \e^{2\phi} F_2^2 H_3^2 + \ldots\Big). \label{la}
\end{align}
Here $\mathcal{R}^2$, $H_3^4$, $H_3^2 \mathcal{R}$, $F_2^4$, etc.~stand collectively for all scalars built from the Riemann tensor and components of $F_2$ and $H_3$ (pulled back to the corresponding brane or O-plane) with arbitrary contractions of tangent or normal indices. The dilaton dependence of the $F_q$ terms arises because the $\alpha^\prime$ expansion of the RR fields is an expansion in powers of $\e^\phi F_q$ rather than $F_q$ alone \cite{Polchinski:1998rr}.
The $c_{ai}$'s are numerical coefficients which can in principle be determined for each type of source by computing string amplitudes or invoking duality arguments (see, e.g., \cite{Bachas:1999um, Wyllard:2000qe, Wyllard:2001ye, Fotopoulos:2001pt, Schnitzer:2002rt, Garousi:2006zh, Garousi:2009dj, Robbins:2014ara, Garousi:2014oya}).
Their values are not important for our analysis and therefore left unspecified in the following.

The dots in \eqref{la} stand for various further terms involving powers of $F_4$, the second fundamental form and/or covariant derivatives of the dilaton and other fields.
Finally, the bulk action \eqref{action} and the Chern-Simons term $\sim \int C_{p_i+1}$ in \eqref{action2} also receive $\alpha^\prime$ corrections. These are again not relevant for the analysis in this paper and therefore not discussed any further.\footnote{As discussed in the introduction, our motivation is to address the issues related to non-standard source terms in the CDT1 and CDT2 models without sacrificing the appealing simplicity of the bulk equations of motion in these models. We will therefore impose that bulk corrections are negligible as in \cite{Cordova:2018dbb, Cordova:2019cvf}. The Chern-Simons action receives corrections involving odd powers of the RR fields (see, e.g., \cite{Garousi:2010rn, Becker:2010ij, Garousi:2011ut, Becker:2011ar, Mashhadi:2020mzf}). If such terms contribute to the scalar potential and other expressions relevant for our no-gos, one can check that they would scale in the same way as the RR-even terms with respect to the relevant field combinations so that we can ignore them for simplicity.}

In this work, we are interested in solutions which preserve maximal symmetry in $4\le d<10$ dimensions. The most general ansatz for the metric is thus
\begin{equation}
\d s_{10}^2 = \e^{2A(y)}g_{\mu\nu}(x)\d x^\mu \d x^\nu + g_{mn}(y) \d y^m \d y^n
\end{equation}
with $\mu,\nu=0,\ldots,d-1$ and $m,n=d,\ldots,9$, where $g_{\mu\nu}$ is the metric of the $d$-dimensional AdS, Minkowski or dS space and $A$ is a function called the warp factor. We will denote the scalar curvatures of $g_{\mu\nu}$ and $g_{mn}$ by $\mathcal{R}_d$ and $\mathcal{R}_{10-d}$, respectively. Maximal symmetry in $d$ dimensions further requires that the dilaton is a function of the internal coordinates only, i.e., $\phi=\phi(y)$, and that $H_3$ and $F_q$ either have zero or (if $q\ge d$) $d$ legs along the $d$-dimensional spacetime.
The equations of motion under the above assumptions are stated in App.~\ref{app:eom0}. In the remainder of this paper, we will set $2\pi\sqrt{\alpha^\prime}=1$ in all expressions.

\subsection{Localized sources and boundary conditions}
\label{sec:bc}

Much of the debate in the earlier works \cite{Cordova:2018dbb, Cribiori:2019clo, Cordova:2019cvf} can be traced back to different assumptions on how to implement the effect of the localized sources in a compactification. Since this will also be important in the present work, let us briefly review the main points of this debate and clarify our perspective on it.

\begin{figure}[t!]
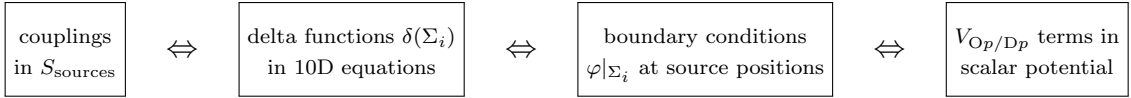

\centering
\begin{equation}
\boxed{\begin{matrix}
{\scriptstyle \text{couplings}} \\[-0.5em]
{\scriptstyle \text{in } S_\text{sources}}
\end{matrix}}
\quad \Leftrightarrow \quad
\boxed{\begin{matrix}
{\scriptstyle \text{delta functions } \delta(\Sigma_i) } \\[-0.5em]
{\scriptstyle \text{in 10D equations}}
\end{matrix}}
\quad \Leftrightarrow \quad
\boxed{\begin{matrix}
{\scriptstyle \text{boundary conditions}} \\[-0.5em]
{\scriptstyle \varphi|_{\Sigma_i} \text{ at source positions}}
\end{matrix}}
\quad \Leftrightarrow \quad
\boxed{\begin{matrix}
{\scriptstyle V_{\text{O}p/\text{D}p} \text{ terms in}} \\[-0.5em]
{\scriptstyle \text{scalar potential}}
\end{matrix}}\,\, \notag
\end{equation}

\caption{Different equivalent ways to describe the effect of localized sources.
\label{schem}}
\end{figure}

The standard way to incorporate D-branes or O-planes in the supergravity equations of motion is to add delta-function sources which are determined by varying the DBI and WZ actions \eqref{action2}.
The delta functions then backreact on the various bulk fields and thus determine their boundary conditions near the sources. Furthermore, the sources contribute to the scalar potential of the compactified theory, as can be verified by dimensionally reducing the 10D action. These different equivalent ways to capture the effect of branes/O-planes are summarized in Fig.~\ref{schem}.

However, as pointed out in \cite{Cordova:2019cvf}, there may be an ambiguity in this prescription.
To see this, we consider the concrete example discussed there, namely an O$8^-$-plane in the CDT1 model \cite{Cordova:2018dbb}. We will study this model in more detail in Section \ref{sec:o8}. For now, we only need to know
the metric $\d s_{10}^2 = \e^{2A(z)} \d s_4^2 + \e^{-2A(z)} \left(\e^{2\lambda(z)} \d s_{\kappa_5}^2 + \d z^2\right)$, where $\kappa_5$ is an Einstein space and $z$ parametrizes a circle (on the covering space of the orientifold). We put the O$8^-$ at $z=0$ and use that $T_{\text{O8}^-}=-32\pi$.
The classical equations of motion (see App.~\ref{app:eom0}) then imply that the warp factor $A$, the dilaton $\phi$ and the conformal factor $\lambda$ satisfy
\begin{align}
(\e^{A-\phi} \phi^{\prime})^\prime &= -20 \delta(z) + \ldots, \label{a1} \\
(\e^{A-\phi} A^{\prime})^\prime &= -4 \delta(z) + \ldots, \label{a2} \\
(\e^{A-\phi} \lambda^{\prime})^\prime &= -8 \delta(z) + \ldots \label{a3}
\end{align}
with $'=\frac{\d}{\d z}$, where we only displayed terms involving a second derivative and the dots stand for various other terms (i.e., flux and curvature terms and products of $A^\prime$, $\phi^\prime$, $\lambda^\prime$). We will ignore the latter since we are only interested in the local field behavior at $z=0$ sourced by the O8$^-$. Here we make the usual assumption that only second derivatives of fields can yield delta functions or, equivalently, that only first derivatives of fields, but not the fields themselves, are allowed to have discontinuities $\sim \text{Heaviside}(z)$.
Let us further assume that the delta functions source the \emph{leading} terms in $A$, $\phi$, $\lambda$ at $z=0$ whereas integrating the dots in \eqref{a1}, \eqref{a2}, \eqref{a3} would only yield higher-order corrections. These assumptions would a priori have to be justified but are often correct and indeed turn out to be self-consistent in the solution of \cite{Cordova:2018dbb, Cordova:2019cvf} considered here.

To solve the equations, it is convenient to combine \eqref{a1} and \eqref{a2} such that
\begin{align}
\Big(\e^{A-\phi}\Big)^{\prime\prime} &= 16\delta(z) + \ldots, \label{a4}
\end{align}
which we can integrate to get $\e^{A-\phi} = c_0 + 8|z| +\ldots $.
Using this in \eqref{a1}, \eqref{a2}, \eqref{a3}, we find
\begin{align}
&c_0\neq 0: && \e^A = a_0 - \frac{2a_0}{c_0} |z|+\ldots, && \e^\phi = \frac{a_0}{c_0} - \frac{10a_0}{c_0^2} |z| +\ldots, && \e^\lambda = l_0 -\frac{4l_0}{c_0} |z|+\ldots, \label{exp1} \\
&c_0=0: && \e^A = a_0 |z|^{-1/4}+\ldots, && \e^\phi = \frac{a_0}{8} |z|^{-5/4}+\ldots, && \e^\lambda = l_0 |z|^{-1/2}+\ldots, \label{exp2}
\end{align}
where $a_0,l_0\neq 0$.
Let us note here that \eqref{a1}, \eqref{a2}, \eqref{a3} are all of the form $f_i^\prime = \delta(z)+ \ldots$ with \emph{finite} $f_i$ (for both $c_0\neq 0$ and $c_0= 0$). Hence, the left-hand-sides of each equation are integrable and well-defined in a distributional sense even though the fields themselves can be divergent. Since we have a local solution, the same must be true for the right-hand sides of the equations, and one can indeed check that this is true. Concerns raised in \cite{Cordova:2019cvf} about having to deal with ill-defined distributions do therefore not apply to our derivation of \eqref{exp1}, \eqref{exp2}.

We can also combine \eqref{a1}, \eqref{a2}, \eqref{a3} such that the delta functions cancel and then divide by $\e^{A-\phi}$. For the solution with $c_0\neq 0$, this yields
\begin{align}
\Big(A-\frac{\phi}{5}\Big)^{\prime\prime} = 0 + \ldots, \qquad \Big(2A-\lambda\Big)^{\prime\prime} = 0 + \ldots, \label{a5}
\end{align}
where the zeros mean that no delta functions are present and the dots stand for all terms without second derivatives as before. One verifies that \eqref{a5} is indeed satisfied by \eqref{exp1} as expected. However, a subtlety arises for the case $c_0=0$. Indeed, it now seems that we can add arbitrary terms $\e^{A-\phi}\delta(z) \sim |z| \delta(z)$ to \eqref{a1}, \eqref{a2} and \eqref{a3} since $|z| \delta(z)=0$ as a distributional product. We thus get
\begin{align}
\Big(A-\frac{\phi}{5}\Big)^{\prime\prime} = C_1 \delta(z) + \ldots, \qquad \Big(2A-\lambda\Big)^{\prime\prime} = C_2 \delta(z) + \ldots, \label{a6}
\end{align}
where $C_1$ and $C_2$ are free parameters.
In other words, the non-associativity of the distributional product $\e^{\phi-A}(\e^{A-\phi}\delta(z)) \neq \delta(z)$ creates an ambiguity in the equations of motion. We again note that $(A-\frac{\phi}{5})^\prime$ and $(2A-\lambda)^\prime$ are finite so that both of the above equations are integrable and well-defined as distributions.
Solving \eqref{a6} yields
\begin{align}
\e^A &= a_0 |z|^{-1/4}+a_1|z|^{3/4}+\ldots, \\ \e^\phi &= \frac{a_0}{8} |z|^{-5/4}+ \frac{5}{16}\left(2a_1- C_1 a_0\right) |z|^{-1/4}+\ldots, \\ \e^\lambda &= l_0 |z|^{-1/2} +\frac{l_0}{2a_0}(4a_1-C_2a_0) |z|^{1/2}+\ldots, &&
\end{align}
whereas this is only a solution of \eqref{a5} for $C_1=C_2=0$. This is precisely the ambiguity in the subleading coefficients observed in \cite{Cordova:2019cvf}, even though we derived it in a different way here.

The boundary conditions with general $a_1$, $C_1$, $C_2$ were called \emph{permissive} in \cite{Cordova:2019cvf}, and we will call those with $C_1=C_2=0$ \emph{classical} in this paper.
Analogously, we will refer to the corresponding delta-function sources in the equations of motion as permissive and classical, respectively. In addition, \cite{Cordova:2019cvf} defined \emph{restrictive} boundary conditions, which aside from $C_1=C_2=0$ also impose $a_1=0$. This is satisfied in simple solutions like an O8 in flat space (see textbooks such as \cite{Ortin:2015hya} for a review), but it is not obvious to us that this needs to be true in general solutions with fluxes and curvature, and we
will therefore not consider the restrictive boundary conditions further.

The crucial question is now whether the correct equations of motion are those with classical sources or those with permissive ones. This is particularly important since the permissive sources lead to dS solutions \cite{Cordova:2018dbb, Cordova:2019cvf, Kim:2020ysx} whereas the classical ones do not \cite{Cribiori:2019clo}.
Here one should distinguish two separate questions, namely whether the permissive sources are allowed in the classical supergravity approximation (which we have assumed in our discussion so far) and whether they are allowed if one takes into account string corrections to the classical supergravity equations.

The first possibility arises due to the ambiguity we just reviewed. This may seem like a technicality with no practical meaning since for $c_0=0$ there is a curvature singularity at $z=0$ which we expect to be cured by string corrections. However, it is in principle possible that these corrections have no other effect than smoothing out the singularity in a small region $|z|\le \epsilon$ so that solving the classical supergravity equations with the included source terms still yields the correct solution for $z> \epsilon$ even in the regime $c_0\to 0$.
It is therefore meaningful to ask whether the permissive sources can make sense already at the level of the classical supergravity equations, and this possibility was indeed advocated in \cite{Cordova:2019cvf}.

However, there are reasons to be skeptical.
In particular, we have seen that the permissive source terms require that the integration constant $c_0$ vanishes. However, $c_0$ is the zero mode of $\e^{A-\phi}$ and therefore a dynamical field, i.e., a modulus. According to \eqref{exp1}, this modulus shifts the dilaton and/or the $\kappa_5$ and circle volumes (depending on whether we keep $a_0$, $l_0$ fixed or not while varying $c_0$). Irrespective of whether this modulus is stabilized or not in a given compactification, it is a degree of freedom which can fluctuate, and such a fluctuation should not create or remove a source. In other words, it should not matter if we first set $c_0=0$ in \eqref{a1}, \eqref{a2}, \eqref{a3} and then derive \eqref{a6} or if we do it the other way round, i.e., first derive \eqref{a6} for finite $c_0$ and then set $c_0=0$. Imposing for consistency that both methods give the same result, we conclude that $C_1=C_2=0$.

This argument would not apply if $c_0$ is for some reason not a dynamical degree of freedom but frozen at $c_0=0$. However, it is not clear to us how to motivate such a proposal. Indeed, the volume and dilaton moduli are believed to be universal in geometric flux compactifications. Such fluctuations also exist when there is non-trivial warping, as has been shown in similar setups (see, e.g., \cite{Giddings:2005ff, Frey:2008xw, Underwood:2010pm}). For the present setup, we will study them in more detail in a separate publication \cite{Junghans:2024}. One might wonder whether the O8 case is special because the corresponding orientifold involution projects out $c_0$, but this is not the case either \cite{Bergshoeff:2001pv}.
An argument given in \cite{Cordova:2019cvf} to nevertheless motivate the permissive boundary conditions was that they are required since otherwise the zero mode of the dilaton has a divergent mass.
However, we will investigate this proposal
in \cite{Junghans:2024} where we find no divergent masses or other pathologies preventing the dilaton from having a dynamical zero mode.

\begin{figure}[t!]
\centering
\includegraphics[trim = 0mm 20mm 0mm 23mm, clip, width=\textwidth]{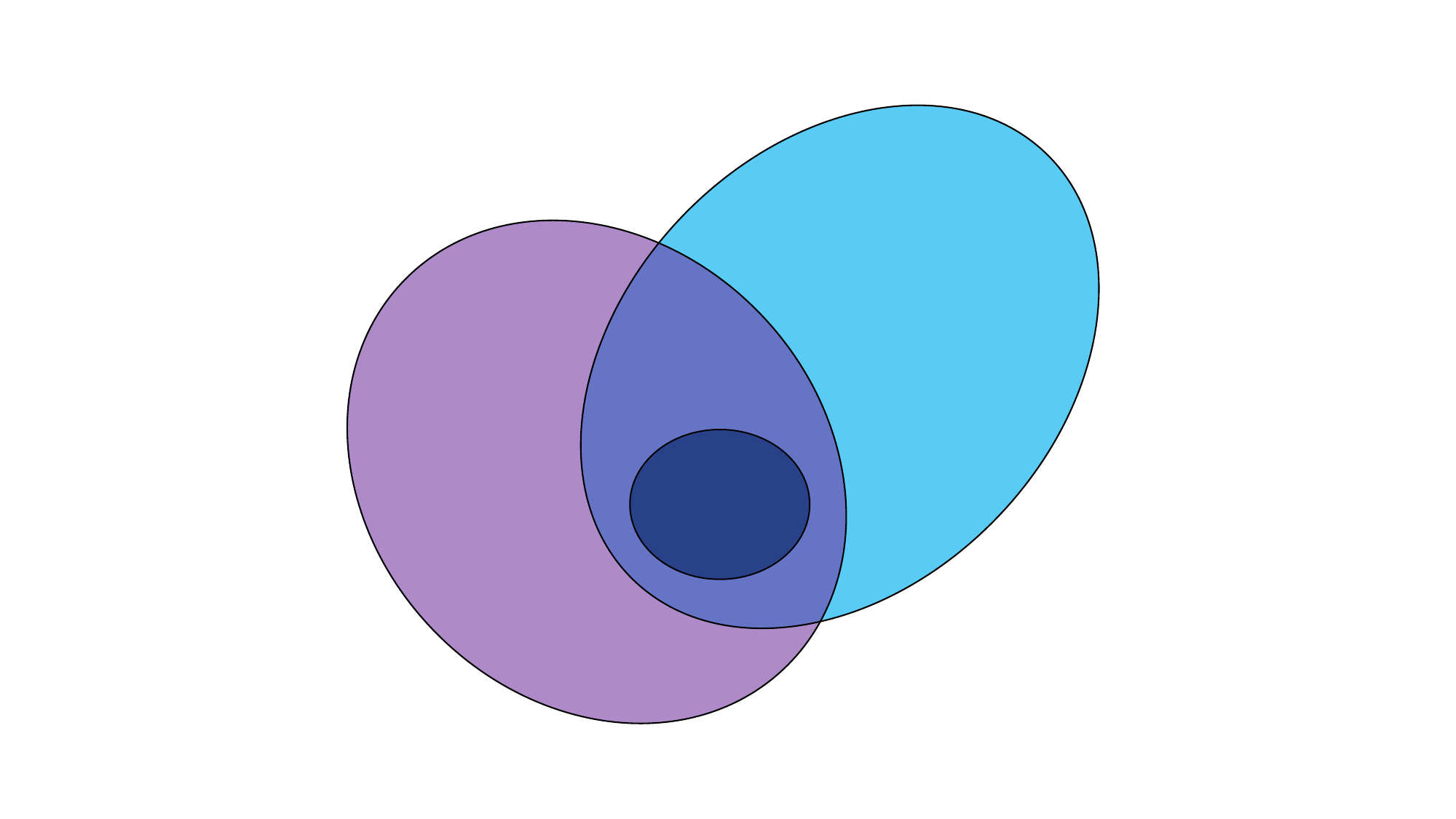}
\put(-247,78){$\scriptstyle{\text{restrictive}}$}
\put(-244,68){$\scriptstyle{C_I=a_1=}$}
\put(-239,58){$\scriptstyle{c_{ai}=0}$}
\put(-252,115){$\scriptstyle{\text{classical}}$}
\put(-255,105){$\scriptstyle{C_I=c_{ai}=0}$}
\put(-321,95){$\scriptstyle{\text{almost}}$}
\put(-323,85){$\scriptstyle{\text{classical}}$}
\put(-319,75){$\scriptstyle{C_I=0}$}
\put(-190,145){$\scriptstyle{\text{permissive}}$}
\put(-183,135){$\scriptstyle{c_{ai}=0}$}

\caption{Different assumptions on the allowed parameter space of O8-plane sources in the CDT1 model.
\label{bc}}
\end{figure}
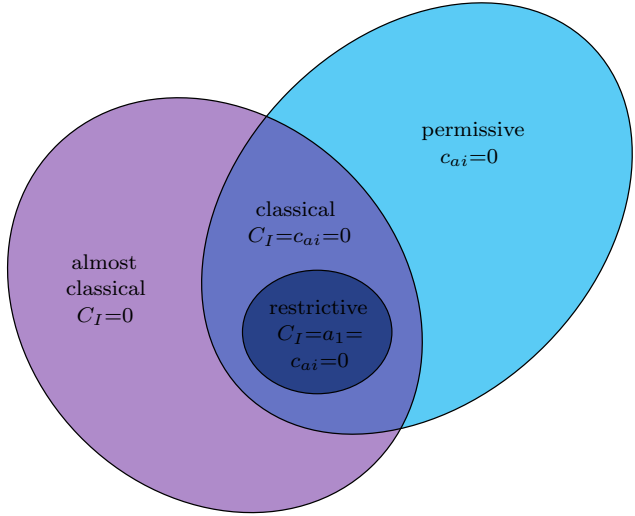

For these reasons, we consider the proposal of permissive source terms implausible at the level of classical supergravity, i.e., one should set $C_1=C_2=0$ in the above equations, consistent with the dS no-go of \cite{Cribiori:2019clo}. On the other hand, as emphasized in \cite{Cribiori:2019clo, Cordova:2019cvf}, it is expected that \emph{string corrections} may in principle change this conclusion. In particular, \cite{Cribiori:2019clo} pointed out that 4-derivative corrections to the classical O-plane/D-brane actions as in \eqref{la} yield new source terms in the equations of motion. We will call such source terms (for arbitrary coefficients $c_{ai}$) \emph{almost classical}, with the classical case corresponding to $c_{ai}=0$. As stated before, the numerical values of these coefficients are fixed in string theory (up to field redefinitions), but we will analyze the effect of the corrections for general $c_{ai}$. See Fig.~\ref{bc} for a summary of the various types of sources/boundary conditions mentioned in this section.

One quickly concludes from \eqref{la} that the field dependence of the 4-derivative corrections is such that the almost classical sources differ somewhat from the permissive ones.
In particular, the 4-derivative terms contribute to the equations of motion even when $\e^A$, $\e^\phi$ and $\e^\lambda$ are finite at $z=0$ instead of abruptly popping up for $c_0=0$ as the permissive sources. Furthermore, the 4-derivative terms generically correct all equations of motion and not just \eqref{a6}.
It is therefore not obvious whether they can lead exactly to the same equations of motion that gave rise to the dS solutions in \cite{Cordova:2018dbb}.
Nevertheless, these solutions demonstrate that allowing source terms other than the classical ones can be enough to obtain dS even in very simple models like the CDT1 model. It is then natural to ask whether the 4-derivative terms predicted by string theory can do the trick as well.
However, we will argue in this paper that this is actually \emph{not} possible in the CDT1 model and variants thereof with O8/D8 sources. We will furthermore find a similar problem for O6-planes in the CDT2 model. Indeed, the dS solutions reported in \cite{Cordova:2019cvf} for that model are again due to non-standard sources but, as we will show, no dS is possible if one considers classical or almost classical ones.

\subsection{The smeared approximation}
\label{sec:smear}

Since parts of our analysis in this paper will use the smeared approximation, we now review the logic behind this approach. Detailed discussions can also be found in \cite[Sec.~4.1]{Junghans:2023lpo}, \cite[Sec.~5]{Junghans:2020acz} and \cite{Baines:2020dmu}.

We start by recalling that the couplings in \eqref{action2} imply that O-planes and D-branes appear as delta-function sources in the equations of motion (cf.~App.~\ref{app:eom0}) and thus backreact on fields such as the warp factor, the dilaton or the internal metric.
This is potentially problematic since the backreaction can create classically singular or strongly curved regions in the vicinity of the sources where supergravity breaks down and string corrections are expected to become important.
To be able to trust supergravity, these stringy regions should vanish or at least be very small, which can be achieved by considering a regime where the backreaction becomes small on most of the spacetime.
In such a regime, the sources only generate negligible spatial variations in the fields such that, e.g., the warp factor and the dilaton are approximately constant. However, this does not necessarily mean that we can simply discard the sources in the equations of motion. Indeed, one is often interested in non-trivial limits where the backreaction becomes small but the tension/charge still gives a relevant contribution, e.g., to the tadpole cancelation condition or the vacuum energy. This can be thought of as expanding the delta functions and the various fields in Fourier modes and considering a limit where only the zero modes are relevant in the equations of motion whereas the effect of all higher modes becomes negligible. The delta functions can thus effectively be replaced by constants integrating to the same value, i.e., the sources are effectively smeared over the transverse space:
\begin{equation}
\delta(\Sigma_i) \to \frac{1}{V_i},
\end{equation}
where $\delta(\Sigma_i)$ is the delta function with support on the worldvolume $\Sigma_i$ (see App.~\ref{app:eom0} for the precise definition) and $V_i$ denotes the volume of the space transverse to the $i$th source. 
Although sometimes claimed in the literature, the existence of such limits is not in conflict with the fact that O-planes are fixed points of an involution and thus localized objects by definition, as can be shown in explicit examples such as the DGKT-CFI AdS vacua \cite{DeWolfe:2005uu, Camara:2005dc, Junghans:2020acz, Marchesano:2020qvg}.

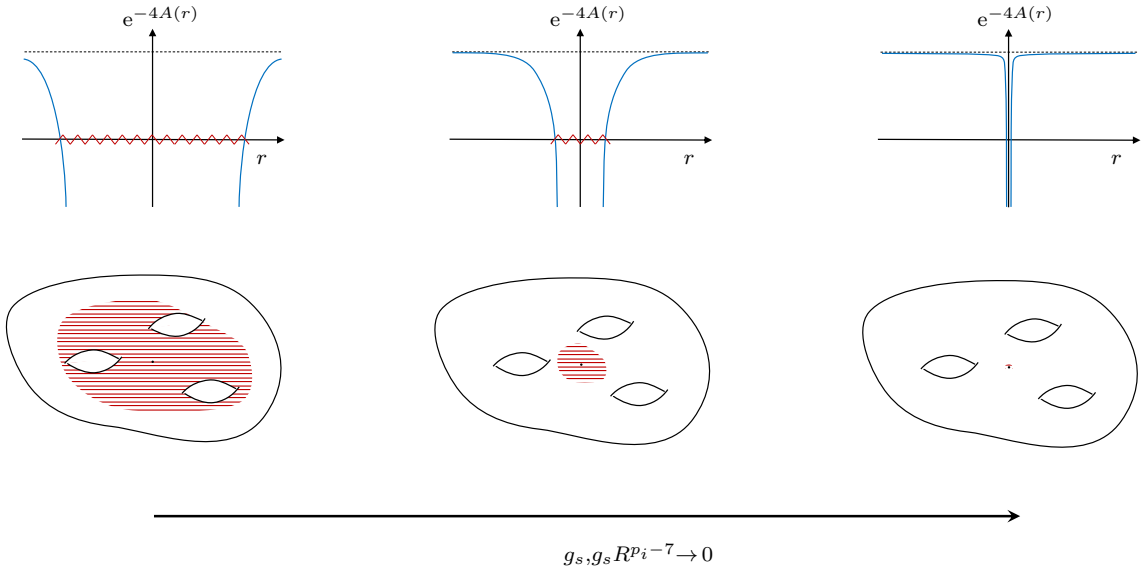
\begin{figure}[t!]
\centering
\includegraphics[trim = 0mm 10mm 0mm 20mm, clip, width=\textwidth]{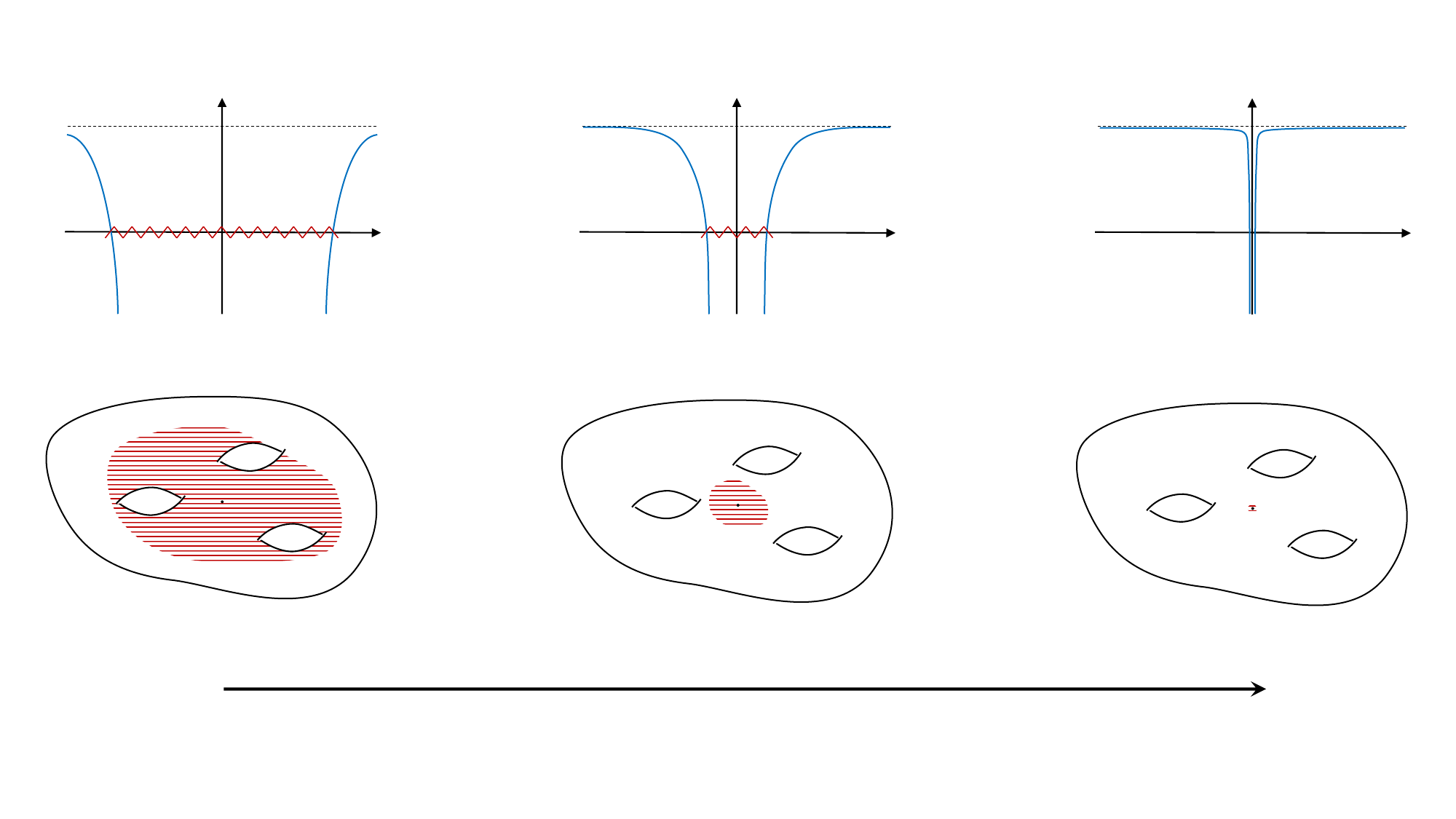}
\put(-230,10){$\scriptstyle{g_s,g_s R^{p_i-7} \to\, 0}$}
\put(-395,212){$\scriptstyle{\e^{-4A(r)}}$}
\put(-235,212){$\scriptstyle{\e^{-4A(r)}}$}
\put(-75,212){$\scriptstyle{\e^{-4A(r)}}$}
\put(-345,160){$\scriptstyle{r}$}
\put(-185,160){$\scriptstyle{r}$}
\put(-25,160){$\scriptstyle{r}$}

\caption{Sketch of the smeared limit on the space transverse to an O$p_i^-$-plane with $p_i\le 6$ (figure adapted from \cite{Junghans:2023lpo}).
For $\mathcal{O}(1)$ volume and string coupling, the O-plane backreacts strongly on a large part of the space (red shade) and fields like the warp factor or the dilaton (blue lines) become ill-defined there, signaling that the classical 10D supergravity description cannot be trusted and string corrections blow up. In the limit $g_s,g_sR^{p_i-7} \to 0$, the backreaction and the stringy region created by it are confined to a smaller and smaller fraction of the transverse space until the field profiles agree with the smeared ones (dashed lines) everywhere.
\label{smeared}}
\end{figure}

A heuristic argument to identify a regime where the backreaction is negligible and smearing is a good approximation goes as follows.
Let us focus on an O$p_i^-$-plane for concreteness and restrict to $p_i\le 6$ for the moment. The backreaction should then be proportional to the tension and the string coupling and fall off like $1/r^{7-p_i}$ with the distance $r$ on the $(9-p_i)$-dimensional transverse space. Hence, it is of the order $g_sT_i/r^{7-p_i}$.
For a space characterized by some length scale $R$, we therefore expect that the backreaction at generic points (i.e., at distances of the order $R$) is of the order
\begin{equation}
\frac{g_sT_i}{R^{7-p_i}}. \label{heur}
\end{equation}
At points very close to the source, the backreaction is larger. In particular,
there is a ball with radius $r \lesssim (g_s|T_i|)^{\frac{1}{7-p_i}}$
in which backreaction effects are $\gtrsim \mathcal{O}(1)$. Fields like the warp factor $\e^{-4A}$ typically become negative and thus ill-defined within this ball, and the curvature diverges at the boundary where the warp factor changes its sign.
These claims can be verified explicitly in the familiar brane solutions in flat space if one chooses a negative tension (see, e.g., \cite{Ortin:2015hya} for a review).
The ball should therefore be thought of as a singular hole in which we cannot trust supergravity and string corrections become relevant, see Fig.~\ref{smeared}.\footnote{To be precise, the region in which we cannot trust supergravity does not start at the hole boundary where the curvature diverges but already at a somewhat larger distance where the curvature becomes $\gtrsim\mathcal{O}(1)$.}

According to the above, the backreaction effects become negligible everywhere in the 10D spacetime in the limit
\begin{equation}
g_s\to 0, \qquad \frac{g_s}{R^{7-p_i}}\to 0. \label{gjlgrhg}
\end{equation}
Indeed, $g_s\to 0$ ensures that the hole
shrinks to a point, and $\frac{g_s}{R^{7-p_i}}\to 0$ ensures that the backreaction goes to zero everywhere else. Since we need large $R$ to control the $\alpha^\prime$ expansion in the bulk, $\frac{g_s}{R^{7-p_i}}\to 0$ is implied by $g_s\to 0$. In this limit, the solution approaches the smeared solution as depicted in Fig.~\ref{smeared}. 
It is therefore plausible\footnote{Strictly speaking, we cannot exclude the possibility of a discontinuous limit such that backreaction effects remain finite in the $d$-dimensional theory when $g_s$ is sent to zero. We assume that this is not the case, which seems physically reasonable.} that the $d$-dimensional effective field theory and, in particular, the scalar potential in this limit are obtained by dimensionally reducing type IIA supergravity in the smeared supergravity background. Assuming that this is correct, it is then clear by continuity that corrections to this scalar potential must be negligible if we slightly move away from the limit to finite but sufficiently small $g_s$ (and finite but sufficiently large $R$).
Hence, for a small hole as on the right in Fig.~\ref{smeared}, we expect that we can use classical supergravity (plus possibly a few leading higher-derivative terms as considered in this paper) to compute the potential and that, at the same time, the smeared approximation is reliable.
In \cite{Junghans:2023lpo}, this was phrased as (see also \cite{Cribiori:2019clo, Junghans:2020acz, Gao:2020xqh} for earlier discussions):
\begin{quote}
{\bf Small-Hole Condition:} \emph{The classical description of a string compactification with O-planes is reliable if the singular/stringy regions generated by the O-planes cover a sufficiently small fraction of the compact space.}
\end{quote}
Whether this can be satisfied in a given model depends on whether there are solutions with sufficiently small $g_s$ and large volume.
To avoid confusions, let us also stress that negligible corrections in $d$ dimensions do not imply negligible corrections at \emph{every} point in the 10D parent theory. Indeed, for any finite $g_s$, string and backreaction corrections remain important in the (possibly very small) hole region surrounding the O-planes in the 10D spacetime. The smeared supergravity solution therefore only gives a reliable description of the low-energy properties of the background but not of the local physics at very small distances from the O-planes.

One may object that the above arguments are all rather heuristic, but they were recently made more precise in \cite{Junghans:2020acz, Marchesano:2020qvg, Junghans:2023lpo}.
These works showed that, at points where the backreaction is small, the 10D solution can be obtained in terms of a perturbative expansion where the leading-order terms correspond to the smeared solution and higher-order corrections encode the backreaction effects.
In particular, \cite{Junghans:2023lpo} showed under fairly general assumptions that the next-to-leading-order backreaction corrections are of the order\footnote{Here we specialize to type IIA string theory in 10 dimensions, i.e., $D=10$ in the notation of \cite{Junghans:2023lpo}.}
\begin{equation}
g_s \sum_i T_{i} G_i, \label{gg}
\end{equation}
where we refer to \cite{Junghans:2023lpo} for the precise numerical coefficients.
Here $G_i$ is the Green's function of the $i$th source satisfying $\nabla^2 G_i = \frac{1}{V_i}-\delta(\Sigma_i)$, where $\nabla^2$ and $V_i$ are the Laplacian and the volume of the $(9-p_i)$-dimensional transverse space with respect to the smeared metric and $\delta(\Sigma_i)$ is the delta function with support on $\Sigma_i$ as before.
In the simple case of an isotropic space with length scale $R$, we have $\nabla^2 \sim R^{-2}$ and $V_i\sim R^{9-p_i}$ and thus the Green's function scales like $G_i \sim 1/R^{7-p_i}$ at generic points. On the other hand, at distances much smaller than $R$, $G_i$ equals the Green's function in flat space up to subleading terms and therefore $G_i\sim 1/r^{7-p_i}$. We thus see that the backreaction is of the order $g_sT_{i}/R^{7-p_i}$ at generic points and $g_sT_{i}/r^{7-p_i}$ for $r\ll R$, which reproduces precisely the heuristic estimates discussed around \eqref{heur} at all points where the backreaction is small.

Importantly, \eqref{gg} remains correct even in cases where simple heuristics fail, in particular in highly anisotropic geometries not characterized by a single length scale $R$. According to \eqref{gg}, the backreaction then vanishes for $g_s \sum_i T_{i} G_i\to 0$. Estimates for the $G_i$'s in such anisotropic limits and the corresponding geometric conditions for small backreaction were discussed in \cite{Cribiori:2021djm, Junghans:2023lpo}.

We further observe that the backreaction starts to blow up for $r \lesssim (g_s|T_{i}|)^{\frac{1}{7-p_i}}$, which is again in agreement with our earlier heuristic argument. Although there is no general proof that the \emph{all}-order backreacted solution must have a singular hole as in Fig.~\ref{smeared} for these values of $r$,
such holes do indeed appear in all solutions where the full backreaction of an O$p_i^-$-plane is known. An intuitive explanation for this is that the near-O$p_i^-$ region in any flux compactification should locally just look like an O$p_i^-$-plane in flat space up to small corrections, which, as already mentioned above, indeed has a singular hole. It therefore seems reasonable to associate strong backreaction with singular holes and demand for control that we are in a regime where the smeared approximation is valid on most of the 10D spacetime.

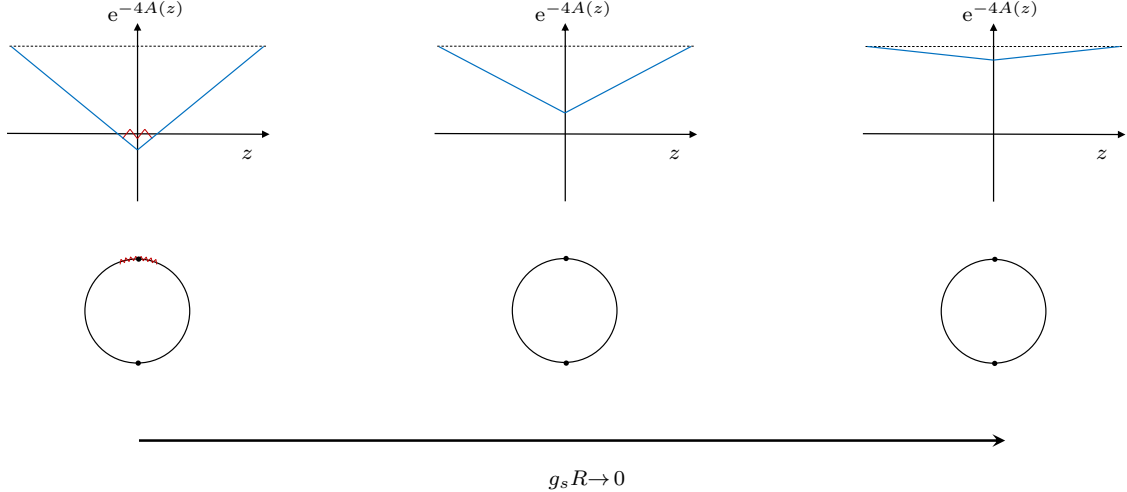
\begin{figure}[t!]
\centering
\includegraphics[trim = 0mm 30mm 0mm 20mm, clip, width=\textwidth]{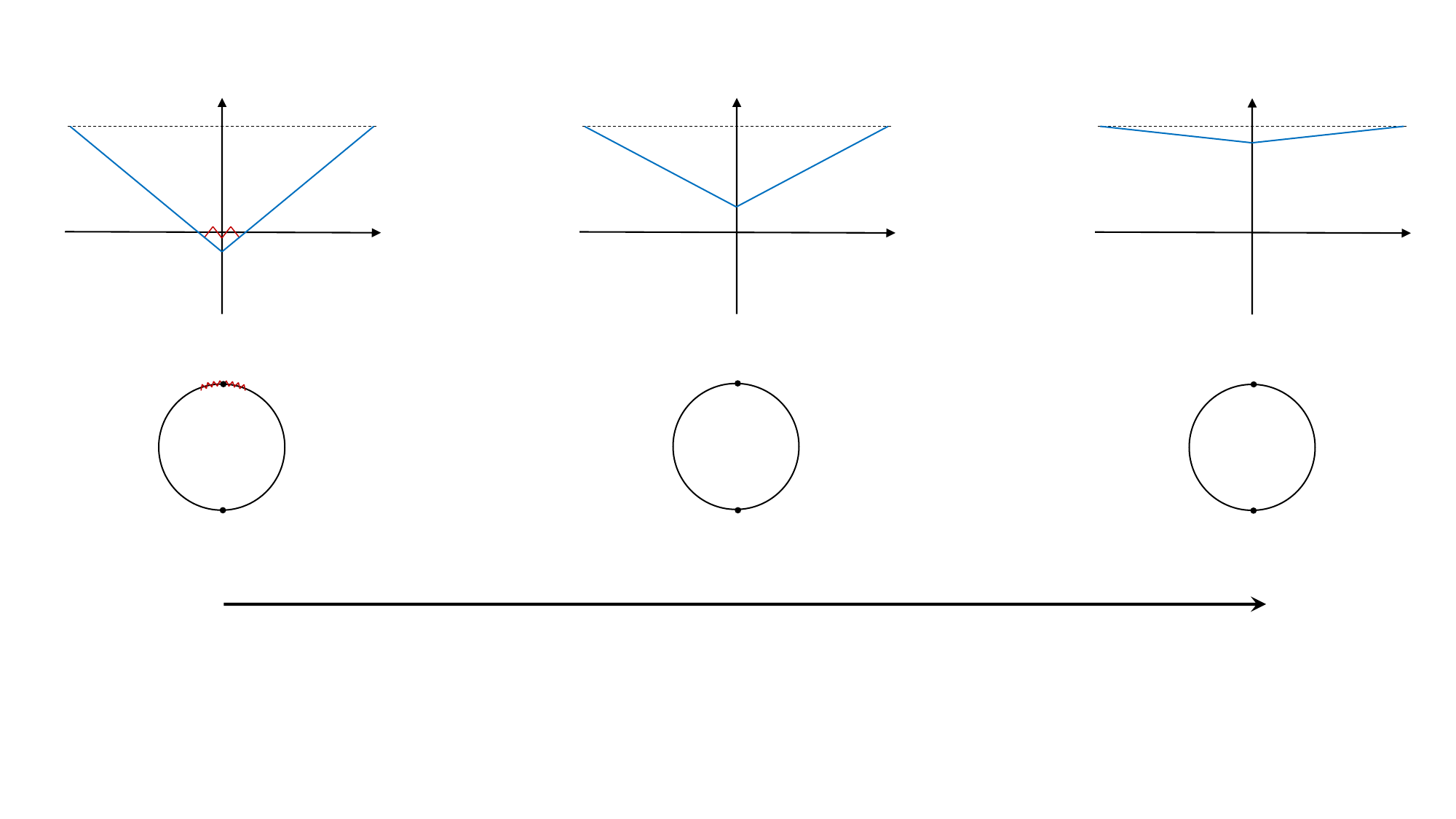}
\put(-230,10){$\scriptstyle{g_s R \to\, 0}$}
\put(-395,185){$\scriptstyle{\e^{-4A(z)}}$}
\put(-235,185){$\scriptstyle{\e^{-4A(z)}}$}
\put(-75,185){$\scriptstyle{\e^{-4A(z)}}$}
\put(-345,133){$\scriptstyle{z}$}
\put(-185,133){$\scriptstyle{z}$}
\put(-25,133){$\scriptstyle{z}$}

\caption{Sketch of the smeared limit for an O$8^-$-plane whose transverse space is a circle with radius $R$ (assuming for concreteness that an O8$^+$-plane at the opposite pole cancels the tadpole).
The smeared limit is approached for $g_sR\to 0$ (right-hand side). For $g_sR\gtrsim \mathcal{O}(1)$, backreaction effects are strong (left and middle) but singular holes need not develop (middle).
\label{smeared2}}
\end{figure}

Let us also briefly comment on positive-tension objects (D-branes, O$^+$-planes). 
In that case, no holes arise in which the warp factor changes its sign. However, for $3<p_i\le 6$, the curvature diverges near the source so that we again have a region where supergravity breaks down and string corrections become important. This stringy region again vanishes in the smeared limit in the same way it does for an O$^-$-plane. However, unlike for O$^-$-planes, there are also other regimes we can consider.
In particular, a stack of $N$ D-branes has small curvature at distances larger than a string length even for $g_sN\gg 1$ (when measured with the supergravity metric) although the warp factor is not approximately constant in this case, as can be verified in the flat-space brane solutions \cite{Ortin:2015hya}.
We therefore expect that solutions with D-branes need not be approximately smeared in order to be reliably described by supergravity.

We finally discuss the case $p_i=8$.
This case is special since
the leading term in the Green's function near the source does not fall off with the distance as for $p_i\le 6$ but is either constant or linear so that some of the above arguments change. The transverse space is one-dimensional and thus, assuming it is compact, a circle with radius $R$. The Green's function is then
\begin{equation}
G = R \left( \frac{z^2}{4\pi}-\frac{|z|}{2}+\frac{\pi}{4} \right) \label{ggg}
\end{equation}
up to an additive constant, where in our convention $z=[0,2\pi)$,
$G$ is zero at the minimum and the source sits at $z=0$.\footnote{The $z^2$ term of \eqref{ggg} cancels out in \eqref{gg} if we include a second source with opposite tension/charge to cancel the tadpole.}
Note that $G$ is of the order $R$ or smaller at every point. According to \eqref{gg}, the backreaction is thus $\lesssim g_s|T_i|R$ for each source and vanishes in the limit $g_sR\to 0$. This is consistent with a small-$g_s$ and large-volume limit as long as $g_s$ goes to zero faster than $R$ goes to infinity. In particular, there are in principle regimes in which the backreaction of $p_i=6$ and $p_i=8$ sources as well as $\alpha^\prime$ and loop corrections are simultaneously negligible.

An important difference to the $p_i\le 6$ case is that the Green's function for $p_i=8$ does not diverge at the source position but stays finite at arbitrarily small distances.
This suggests that the warp factor and the other fields can stay finite as we approach the source in the all-order backreacted solution, which is indeed true in the known non-compact O$8^-$ solution \cite{Ortin:2015hya}.
We thus conclude that the backreaction of an O$8^-$-plane does not necessarily create a singular hole or large curvature, see Fig.~\ref{smeared2}.
Similar remarks apply to D8-branes and O$8^+$-planes in a compact setting.

Let us summarize our discussion in this section. We argued that spacetime regions in which O$p_i^-$-planes with $p_i\le 6$ backreact strongly have singular holes in which classical supergravity breaks down and string corrections are important. We further explained that, at sufficiently small $g_s$ and large $R$, these holes become negligibly small and the solution is well-approximated by the smeared one on most of the 10D spacetime.
On the other hand, for O$8^\pm$-planes and D-branes, having a strong backreaction does not imply a singular hole or large curvature.
Solutions involving such sources can therefore be reliable in supergravity even if they are not approximately smeared.

In the remainder of this paper, we will often assume the existence of a regime where the smeared approximation is valid, as this guarantees a good control over the dangerous holes and their associated string corrections. However, we will also see that at least some of our arguments for smeared solutions have fully backreacted analogues which are in full agreement with our smeared results.

\section{Models with $p_i=8$ sources}

\label{sec:o8}

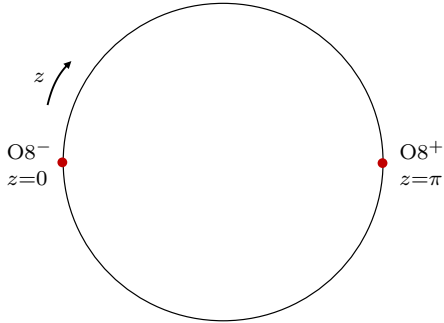
\begin{figure}[t!]
\centering
\includegraphics[trim = 0mm 50mm 0mm 40mm, clip, width=\textwidth]{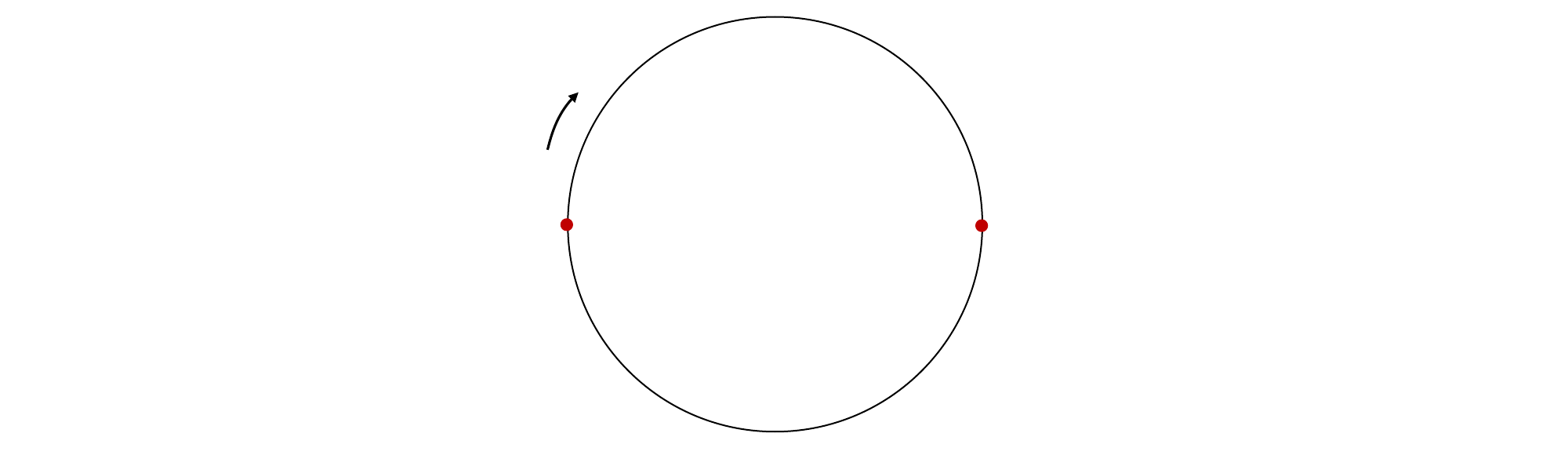}
\put(-300,100){$\scriptstyle{z}$}
\put(-310,71){$\scriptstyle{\text{O}8^-}$}
\put(-310,61){$\scriptstyle{z=0}$}
\put(-163,71){$\scriptstyle{\text{O}8^+}$}
\put(-163,61){$\scriptstyle{z=\pi}$}

\caption{A circle parametrized by $z\in[0,2\pi)$ with an O$8^-$-plane at $z=0$ and an O$8^+$-plane at $z=\pi$.
\label{s1}}
\end{figure}

In this section, we study models with $p_i=8$ sources (i.e., O$8^\pm$-planes and D8-branes) and $F_0$ as the only flux. This includes in particular the CDT1 model which was argued in \cite{Cordova:2018dbb} to have dS solutions. The internal space in this model is a negatively curved Einstein manifold times a circle, and the orientifold involution yields an O$8^-$-plane and a parallel O$8^+$-plane which wrap the Einstein manifold and are pointlike on the circle, see Fig.~\ref{s1}.\footnote{\cite{Cordova:2018dbb} also discusses a variant of the model involving $F_4$ flux which will not be studied here.}
However, our arguments in this section apply more generally as they do not depend on the source types (O-planes or D-branes), their distribution, the dimension $d$ and the space we compactify on.

It was already shown in \cite{Cribiori:2019clo} that models of the above kind cannot have dS solutions \emph{classically} since the equations of motion imply a vanishing vacuum energy if only the leading terms in the effective action are taken into account (i.e., \eqref{action} and \eqref{action2} with $\mathcal{L}_{\alpha^{\prime 2},i}=0$). We will review this argument below, both from the point of view originally discussed in \cite{Cribiori:2019clo} considering the fully backreacted 10D solution and from the point of view of the scalar potential of the $d$-dimensional effective field theory obtained in the smeared approximation, which provides an alternative perspective but leads to the same conclusions.

As explained in \cite{Cribiori:2019clo, Cordova:2019cvf} and in Section \ref{sec:bc}, the dS solutions of \cite{Cordova:2018dbb} formally avoid the classical no-go by having non-standard source terms in the equations of motion.
It is therefore natural to wonder whether $\alpha^\prime$ corrections to the O8/D8 actions as in \eqref{la} can have the same effect of allowing dS vacua. This is the main question we seek to answer in this section.\footnote{A separate issue pointed out in \cite{Cordova:2018dbb, Bena:2020qpa} is that dS solutions in the CDT1 model may have a brane instability if its O$8^+$-plane is in reality an O$8^-$-plane with 32 D8-branes on top.}

\subsection{Scalar potential}
\label{sec:scal}

We first compute the $d$-dimensional scalar potential. To this end, we perform a dimensional reduction of the type IIA supergravity action down to $d$ dimensions (as, e.g., in \cite{Hertzberg:2007wc}).
For simplicity, we consider the smeared approximation. As explained in Section \ref{sec:smear}, this is motivated by the expectation that the existence of an approximately smeared regime is sufficient (although not always necessary) to ensure that no large singular holes appear and classical supergravity is reliable.

It will be sufficient for our arguments to keep two moduli, the overall volume and the dilaton. We therefore redefine $g_{mn}=\rho^{\frac{2}{10-d}}\hat{g}_{mn}$ such that $\hat{g}$ has unit volume and $\rho=\int \d^{10-d} y \sqrt{g_{10-d}}$ denotes the volume modulus. Furthermore, we define $\tau=\e^{-\phi}$ as the dilaton modulus. In general, the scalar potential depends on many other moduli as well (e.g., cycle volumes and axions) but we keep this dependence implicit as it is not relevant for our analysis. 

Since $H_3=F_q=0$ for $q\neq 0$ in the models we consider, the possible 4-derivative terms are of the form
\begin{equation}
\mathcal{L}_{\alpha^{\prime 2},i} = \left(c_{1i} (\e^\phi F_0)^4 + c_{2i} (\e^\phi F_0)^2 \mathcal{R} + c_{3i} \mathcal{R}^2 \right).
\end{equation}
We remind the reader that by $\mathcal{R}^2$ we do not necessarily mean the square of the Ricci scalar but any contraction of two Riemann tensors (or a sum thereof) with tangent or normal indices with respect to the corresponding O8-plane or D8-brane.
In practice, only terms involving internal components of the Riemann tensor will be relevant for us, as they correct the scalar potential. On the other hand, external Riemann tensors yield either curvature-squared terms or a correction to the Einstein-Hilbert term in the $d$-dimensional theory, which we can ignore in the relevant regime of small curvature.\footnote{We can furthermore ignore terms involving the second fundamental form or covariant derivatives since they would have the same scaling with the field $\beta$ defined below as the displayed terms and would therefore not change our conclusions. For the same reason, we do not consider the possibility of Chern-Simons corrections involving odd powers of $\e^\phi F_0$.}

Dimensionally reducing the type IIA supergravity action given by \eqref{action}, \eqref{action2}, we obtain\footnote{Reducing the RR term of the democratic action is subtle due to the duality constraint and effectively leads to a doubling of the numerical coefficient, as explained, e.g., in \cite{DeWolfe:2002nn} for the case of $F_5$ in IIB.}
\begin{align}
S \supset 2\pi \int \d^d x \sqrt{-g_d} & \Bigg(\tau^2 \rho  \mathcal{R}_d + \tau^2 \rho^{\frac{8-d}{10-d}} \hat{\mathcal{R}}_{10-d} - \frac{1}{2} \rho F_0^2
-\tau \rho^{\frac{9-d}{10-d}} \sum_i \frac{T_i}{2\pi \hat V_i} \Big(1+c_{1i} \tau^{-4} F_0^4 \nl + c_{2i} \tau^{-2}\rho^{-\frac{2}{10-d}} F_0^2 \hat{\mathcal{R}}_{10-d} + c_{3i} \rho^{-\frac{4}{10-d}} \hat{\mathcal{R}}^2_{10-d} \Big)\Bigg),
\end{align}
where $i$ runs over the O$8^\pm$-planes and D8-branes and the hats indicate that the corresponding objects are defined with respect to $\hat g_{mn}$.

We now go to the $d$-dimensional Einstein frame by performing the further redefinition
\begin{equation}
g_{\mu \nu}=\tau^{-\frac{4}{d-2}} \rho^{-\frac{2}{d-2}} g_{\mu \nu}^E.
\end{equation}
We thus find
\begin{equation}
S \supset 2 \pi \int \d^d x \sqrt{-g_d^E} \left(\mathcal{R}_d^E - V\right)
\end{equation}
with the scalar potential
\begin{align}
V&=-\tau^{-\frac{4}{d-2}} \rho^{-\frac{16}{(10-d)(d-2)}} \hat{\mathcal{R}}_{10-d} + \frac{1}{2} \tau^{-\frac{2d}{d-2}} \rho^{-\frac{2}{d-2}} F_0^2 \nl
+ \tau^{-\frac{d+2}{d-2}} \rho^{-\frac{18-d}{(10-d)(d-2)}} \sum_i \frac{T_i}{2 \pi \hat V_i} \Big(1+c_{1i} \tau^{-4} F_0^4
+ c_{2i} \tau^{-2} \rho^{-\frac{2}{10-d}} F_0^2 \hat{\mathcal{R}}_{10-d} \nl +c_{3i} \rho^{-\frac{4}{10-d}} \hat{\mathcal{R}}^2_{10-d} \Big).
\end{align}
Note that the $d$-dimensional traced Einstein equation is
\begin{equation}
\mathcal{R}_d^E=\frac{d}{d-2}V, \label{deom}
\end{equation}
and therefore the vacuum energy is determined by the on-shell scalar potential.

To analyze the extrema of $V$, it is convenient to make another field redefinition
\begin{equation}
\rho=\left(\alpha \beta\right)^{10-d},\qquad \tau=\beta. \label{rt}
\end{equation}
We also introduce the notation
\begin{equation}
T \equiv \sum_i \frac{T_i}{2 \pi \hat V_i}.
\end{equation}
This yields
\begin{align}
V &= \beta^{-\frac{20}{d-2}} \left(-\alpha^{-\frac{16}{d-2}} \hat{\mathcal{R}}_{10-d} + \frac{1}{2} \alpha^{-\frac{2(10-d)}{d-2}} F_0^2 + \alpha^{-\frac{18-d}{d-2}} T \right) \nl
+ \beta^{-\frac{4(d+3)}{d-2}} \alpha^{-\frac{18-d}{d-2}} \sum_i \frac{T_i}{2 \pi \hat V_i} \left( c_{1i} F_0^4 + c_{2i} \alpha^{-2} F_0^2 \hat{\mathcal{R}}_{10-d} + c_{3i} \alpha^{-4} \hat{\mathcal{R}}^2_{10-d} \right). \label{pot}
\end{align}

Depending on the type of model, the potential further simplifies.
In particular, the $F_0$ terms must vanish in models which involve an orientifolding since $F_0$ is odd under the involution in the $p_i=8$ case \cite{Bergshoeff:2001pv} and constant in the smeared solution, which together implies $F_0=0$. This is consistent with the fact that, in the backreacted solution of \cite{Cordova:2018dbb}, $F_0$ flips its sign as it crosses one of the O8-planes, implying a vanishing average/zero mode. Hence, $F_0$ is purely a backreaction effect here\footnote{Characterizing $F_0$ as a backreaction effect may be confusing since it is piecewise constant in the solution of \cite{Cordova:2018dbb} and so does not fall off with the distance from the sources. However, note that the backreaction of a source is only expected to fall off for codimension $\ge 3$ whereas here the codimension is 1.} (and thus negligible in $V$ by assumption of the smeared limit, cf.~Section \ref{sec:smear}), in analogy to how $F_2$ in compactifications with O6-planes is generated by the backreaction even when there is no topological $F_2$ flux \cite{Blaback:2010sj, Junghans:2020acz, Marchesano:2020qvg}.
The potential in models containing O8-planes thus simplifies to
\begin{align}
V &= \beta^{-\frac{20}{d-2}} \left(-\alpha^{-\frac{16}{d-2}} \hat{\mathcal{R}}_{10-d} + \alpha^{-\frac{18-d}{d-2}} T \right) 
+ \beta^{-\frac{4(d+3)}{d-2}} \alpha^{-\frac{10+3d}{d-2}} \sum_i \frac{T_i}{2 \pi \hat V_i} c_{3i} \hat{\mathcal{R}}^2_{10-d}. \label{pot1b}
\end{align}
On the other hand, in models which do not contain O8-planes but only D8-branes, $F_0$ need not vanish so that the more general potential \eqref{pot} applies. However, note that the only way to cancel the tadpole created by the D8-branes is then to add anti-D8-branes, which is expected to yield an open-string tachyon. It is therefore not clear to us whether there are situations where the general potential \eqref{pot} with $F_0$ terms turned on is relevant, at least when we are interested in vacua rather than running solutions.

A further simplification in the CDT1 model of \cite{Cordova:2018dbb} is that $T$ vanishes since $T_{\text{O}8^-}=-T_{\text{O}8^+}$ and $\hat V_i$ is the same circle volume for both O-planes. Hence, the second term on the right-hand side in \eqref{pot1b} vanishes.
Moreover, if the coefficient $c_{3i}$ of the $\alpha^\prime$ correction is the same for O$8^-$-planes and O$8^+$-planes, the third term in \eqref{pot1b} is proportional to $T$ and thus vanishes as well. In that case, the potential is uncorrected at the 4-derivative level and simply equals the classical one.
The same conclusions would hold in models where the tadpole of the O$8^-$-plane is cancelled by D8-branes instead of an O$8^+$-plane since such models again have $T=0$. However, $T$ can be non-zero in models which include anti-D8-branes and/or anti-O8$^\pm$-planes since in that case tadpole cancelation does not imply that the tensions must also cancel.

In the following, we will
study the general potential \eqref{pot} without making any assumptions about $F_0$ or $T$. Interestingly, we will find that our main conclusion does not depend on whether these terms have to be set to zero in a specific model. In particular, we will show that meta-stable dS vacua are ruled out, both classically and including the $\alpha^\prime$ corrections.

\subsection{Classical no-go}
\label{sec:o8nogo}

In this section, we show that dS is ruled out classically in the CDT1 model and other compactifications with O8/D8 sources and $F_0$ flux. We will first review the no-go of \cite{Cribiori:2019clo}, which is based on the 10D equations of motion and takes into account the full backreaction of the sources.\footnote{See also \cite{Andriot:2016xvq} for a similar no-go assuming a less general ansatz.} We will then show how to reproduce the same result from the perspective of the smeared scalar potential we just computed.

The starting point of the no-go of \cite{Cribiori:2019clo} is to combine the 10D Einstein and dilaton equations such that the O8/D8 source terms and the $F_0^2$ energy density cancel out. This yields\footnote{Note that \cite{Cribiori:2019clo} used the Einstein frame whereas here we use the string frame. Furthermore, the no-go  in \cite{Cribiori:2019clo} was derived for any type of $F_q$ and $H_3$ flux while our focus is on the special case with only $F_0$.}
\begin{align}
\e^{-2A} \mathcal{R}_d &= \frac{\e^{-dA+2\phi}}{\sqrt{g_{10-d}}} \partial_m\left( \sqrt{g_{10-d}}\, \e^{\frac{d-10}{5}\phi}\, \partial^m \e^{dA-\frac{d}{5}\phi} \right),
\label{r}
\end{align}
where we
assumed the classical equations of motion (i.e., those arising from the variation of \eqref{action}, \eqref{action2} with $\mathcal{L}_{\alpha^{\prime 2},i}=0$ as stated in App.~\ref{app:eom0}) with $p_i=8$ sources and $F_0$ flux. Multiplying both sides of \eqref{r} by $\e^{dA-2\phi}$ and integrating over the compact space, one finds
\begin{equation}
\left(\int\d^{10-d} y\sqrt{g_{10-d}} \,\e^{(d-2)A-2\phi}\right) \mathcal{R}_d = 0. \label{sggsghsg}
\end{equation}
We thus see that only Minkowski vacua are possible classically.\footnote{If one turns on further fluxes such as $F_4$, the classical vacuum energy is negative \cite{Cribiori:2019clo}.}

How should we interpret this result? It is clear that \eqref{sggsghsg} does not imply an exactly vanishing vacuum energy since it ignores string corrections.
As emphasized in \cite{Cribiori:2019clo}, the lesson to be learned from \eqref{sggsghsg} is rather that the contribution of the string corrections to the vacuum energy is always \emph{leading} since the classical contribution vanishes. A classical computation can therefore never determine the sign of the vacuum energy in the CDT1 model and similar O8/D8 models, regardless of how small $g_s$ and how large the volume are.

Assuming small enough $g_s$ and large enough volume that curvatures are weak everywhere on the 10D spacetime (as in the middle and on the right in Fig.~\ref{smeared2}), the leading corrections to \eqref{sggsghsg} come from the next-to-leading-order terms in the $\alpha^\prime$ expansion in the Einstein and dilaton equations, i.e., either from 8-derivative corrections in the bulk or localized 4-derivative corrections to the O8/D8 source terms. The latter generically dominate over the bulk terms and thus decide the sign of $\mathcal{R}_d$ unless there are specific cancelations.

If $g_s$ exceeds a critical value, holes begin to develop around the O$8^-$-plane(s) in which the $\alpha^\prime$ expansion breaks down (as on the left in Fig.~\ref{smeared2}). One may then be worried that integrating \eqref{r} over the internal space does not make sense anymore, as emphasized in \cite{Cordova:2019cvf}. To see that this does not change the conclusion, we can integrate \eqref{r} up to a boundary which is chosen such that it cuts out the unreliable hole regions around each O$8^-$. This creates boundary terms $\mathcal{B}_i(\epsilon)$ in \eqref{sggsghsg} whose values depend on the hole diameter $\epsilon$. Let us define $g_{s,\text{crit}}$ as the largest value of $g_s$ for which the $\alpha^\prime$ expansion is still ok everywhere, i.e., we have $\epsilon\to 0$ for $g_s\to g_{s,\text{crit}}$. Then, in the vanishing-hole limit, each $\mathcal{B}_i(0)$ term must match with the 4-derivative (and higher) corrections from the corresponding O$8^-$ that sits inside the hole. In particular, as shown above, there is no contribution to \eqref{sggsghsg} from the classical tension of the O$8^-$ in this limit, and therefore the lowest order at which O$8^-$ terms can contribute is at 4 derivatives. For finite but small hole diameter (corresponding to a small increase of $g_s$ beyond $g_{s,\text{crit}}$), the boundary terms are approximately the same as before, $\mathcal{B}_i(\epsilon)=\mathcal{B}_i(0)+\mathcal{O}(\epsilon)$, and therefore the conclusions drawn for $\epsilon=0$ still apply, i.e., each hole correction to \eqref{sggsghsg} approximately equals a localized 4-or-higher-derivative term evaluated at $g_s=g_{s,\text{crit}}$.
Therefore, the central claim of \cite{Cribiori:2019clo} that the sign of the vacuum energy is undetermined \emph{classically} and can only be fixed by computing $\alpha^\prime$ corrections remains correct even in the presence of holes, as long as they are sufficiently small.
On the other hand, for large holes, we expect that the supergravity calculations in this paper and in \cite{Cordova:2018dbb, Cribiori:2019clo, Cordova:2019cvf} cannot be trusted and one should instead use non-perturbative methods to study string vacua in such regimes.

Since it will be relevant for the rest of this paper, let us also reproduce our arguments from the point of view of the scalar potential of the $d$-dimensional effective field theory. The classical potential (in the smeared approximation) is obtained from \eqref{pot} by setting the $\alpha^{\prime}$ corrections on the O8-planes/D8-branes to zero ($c_{ai}=0$). This yields
\begin{align}
V_\text{class}
&= \beta^{-\frac{20}{d-2}} \left(-\alpha^{-\frac{16}{d-2}} \hat{\mathcal{R}}_{10-d} + \frac{1}{2} \alpha^{-\frac{2(10-d)}{d-2}} F_0^2 + \alpha^{-\frac{18-d}{d-2}} T \right).
\end{align}
Using the equation of motion $\beta\partial_\beta V_\text{class} = -\frac{20}{d-2} V_\text{class} = 0$, it follows that the classical vacuum energy vanishes,
\begin{equation}
V_\text{class}|_0 = 0, \label{rr}
\end{equation}
where $|_0$ stands for on-shell evaluation. According to \eqref{deom}, this implies a classical Minkowski vacuum. Hence, as expected, the $d$-dimensional argument is in agreement with the 10D argument leading to \eqref{sggsghsg}.

The full scalar potential $V$ is
\begin{equation}
V = V_\text{class} + \delta V_\text{stringy}, \label{rrr}
\end{equation}
where $\delta V_\text{stringy}$ denotes all possible string corrections to $V_\text{class}$, both from the bulk spacetime and the hole regions/O-plane sources.
In order to trust a classical calculation of the (AdS or dS) vacuum energy, all string corrections we ignore must be subleading, i.e., we require
\begin{equation}
V_\text{class}|_0\neq 0 \qquad \text{and} \qquad V_\text{class}|_0 \gg \delta V_\text{stringy}|_0. \label{crit}
\end{equation}
Similarly, the moduli masses should satisfy $(m^2)_\text{class}\neq 0$ and $(m^2)_\text{class}\gg \delta(m^2)_\text{stringy}$ or else a classical calculation is not sufficient to predict the stability of the solution.

As an explicit example meeting these criteria, consider the DGKT-CFI class of AdS flux vacua in type IIA string theory \cite{DeWolfe:2005uu, Camara:2005dc}.
While it is not yet rigorously proven that these vacua are non-perturbatively consistent, they do pass the above sanity check of a self-consistent approximation scheme: Indeed, one can argue that \eqref{crit} holds at large 4-form flux for these vacua \cite[Sec.~5]{Junghans:2020acz} such that
\begin{equation}
V|_0 \approx V_\text{class}|_0 \label{dlsjdjlsdg}
\end{equation}
up to parametrically small corrections.

One might have hoped that the CDT1 model or similar O8/D8 models have dS vacua which are classical in the same way, i.e., satisfy $V_\text{class} |_0 >0$ with negligible $\delta V_\text{stringy}|_0$.
However, the no-go of \cite{Cribiori:2019clo} shows that this is ruled out and instead, according to \eqref{sggsghsg} and \eqref{rr},
\begin{equation}
V |_0 = \delta V_\text{stringy}|_0. \label{sliglsjg}
\end{equation}
Hence, as we already observed from the 10D perspective, the leading vacuum energy is generated by string corrections.
Although this does not exclude the possibility of dS vacua per se, it does imply that this question is undecidable by a classical calculation.
We stress once more that this did not have to be the case: although the \emph{short-distance} physics of an O-plane with a hole or a strongly curved region is non-perturbative and can therefore as a matter of principle not be understood using classical supergravity, it could have been that classical supergravity is sufficient to compute the leading contribution to the \emph{vacuum energy}, as in other compactifications with O-planes.

In the concrete example of the CDT1 model, \cite{Cordova:2018dbb} found that the classical equations of motion can be solved numerically when imposing $\mathcal{R}_4>0$ and interpreted this as evidence for dS vacua (see also \cite{Kim:2020ysx} for the corresponding analytic solution).
However, according to \eqref{sliglsjg}, this corresponds to implicitly assuming $\delta V_\text{stringy}|_0 >0$, and since the sign of $\delta V_\text{stringy}|_0$ was not determined in \cite{Cordova:2018dbb}, it is a priori equally plausible that $\delta V_\text{stringy} |_0\le 0$ is realized in string theory, which would mean that the numerics have to be run imposing zero or negative $\mathcal{R}_4$. It was therefore concluded in \cite{Cribiori:2019clo} that the classical calculation of \cite{Cordova:2018dbb} does by itself not provide evidence favoring dS over AdS/Minkowski.

It was also pointed out in \cite{Cribiori:2019clo} that it might be possible to avoid the classical dS no-go by taking into account string corrections in a controlled way. In particular, a relatively mild modification of the classical-dS scenario is to incorporate the leading $\alpha^\prime$ corrections to the O8/D8 actions at the 4-derivative order.
As explained above, such 4-derivative terms may in principle violate the no-go and provide the leading contribution to the vacuum energy.
Furthermore, including these corrections is motivated by the observation in \cite{Cribiori:2019clo} (see also \cite{Cordova:2019cvf}) that the dS solution of \cite{Cordova:2018dbb} avoids the no-go in spite of solving the classical bulk equations by having non-standard O8 boundary conditions which are incompatible with the classical source terms (cf.~the discussion in Section \ref{sec:bc}).
A reasonable hope is therefore that considering the effect of the 4-derivative terms is sufficient to lift the classically vanishing vacuum energy to a positive value with $\delta V_\text{stringy}|_0>0$ while further string corrections beyond the 4-derivative ones are still self-consistently suppressed in the potential. We will analyze this almost-classical-dS scenario in the next subsection.

\subsection{Adding $\alpha^\prime$ corrections}
\label{sec:o8nogo2}

Recall from Section \ref{sec:scal} that the $\alpha^\prime$-corrected scalar potential up to the 4-derivative level is
\begin{align}
V &= \beta^{-\frac{20}{d-2}} \left(-\alpha^{-\frac{16}{d-2}} \hat{\mathcal{R}}_{10-d} + \frac{1}{2} \alpha^{-\frac{2(10-d)}{d-2}} F_0^2 + \alpha^{-\frac{18-d}{d-2}} T \right) \nl
+ \beta^{-\frac{4(d+3)}{d-2}} \alpha^{-\frac{18-d}{d-2}} \sum_i \frac{T_i}{2 \pi \hat V_i} \left( c_{1i} F_0^4 + c_{2i} \alpha^{-2} F_0^2 \hat{\mathcal{R}}_{10-d} + c_{3i} \alpha^{-4} \hat{\mathcal{R}}^2_{10-d} \right) \label{pot2}.
\end{align}
Classically, the second line vanishes and $V \sim \beta^{-\frac{20}{d-2}}$ has a runaway behavior unless the condition
\begin{equation}
\label{NoScaleCond}
-\alpha^{-\frac{16}{d-2}} \hat{\mathcal{R}}_{10-d}+ \frac{1}{2} \alpha^{-\frac{2(10-d)}{d-2}} F_0^2 + \alpha^{-\frac{18-d}{d-2}} T=0
\end{equation}
is satisfied at the extremum for $\alpha$.\footnote{For brevity, we omit the subscript $|_0$ for on-shell quantities from now on.}
In the latter case, $V=0$ and $\beta$ is a flat direction, in agreement with the no-go theorem we reviewed above.
This is reminiscent of the GKP solution in type IIB string theory \cite{Giddings:2001yu} where the scalar potential has a runaway behavior for the volume unless one imposes a  condition similar to \eqref{NoScaleCond} for the 3-form fluxes. In the latter case, $V=0$ in GKP and the volume is a flat direction.

Including the $\alpha^\prime$ corrections in the second line in $V$, we can stabilize $\beta$ if the remaining moduli are stabilized such that \eqref{NoScaleCond} is violated (otherwise we again get a runaway potential $V\sim \beta^{-\frac{4(d+3)}{d-2}}$), i.e., we require
\begin{equation}
\label{NoScaleCond2}
-\alpha^{-\frac{16}{d-2}} \hat{\mathcal{R}}_{10-d}+ \frac{1}{2} \alpha^{-\frac{2(10-d)}{d-2}} F_0^2 + \alpha^{-\frac{18-d}{d-2}} T \equiv C \neq 0.
\end{equation}
Stabilizing $\beta$ this way means that we balance the $\alpha^\prime$ corrections with the leading terms in \eqref{pot2}. In the perturbative regime of small curvature and energy densities, this is typically only possible if the sum \eqref{NoScaleCond2} is fine-tuned to be much smaller than each of its three terms individually.

We will not perform a detailed analysis of whether such a fine-tuning is possible in concrete models since the above potential has in fact a more serious and immediate problem. In particular, using $\partial_\beta V=0$, one verifies that on-shell
\begin{equation}
V= - \frac{(d-2)^2}{80(d+3)} \beta^2\partial_\beta^2 V. \label{o8nogo3}
\end{equation}
This implies
\begin{equation}
\partial_\beta^2 V <0 \label{lkthkth}
\end{equation}
whenever the vacuum energy is positive. By Sylvester's criterion, the Hessian then has a negative eigenvalue and we get a tachyon.

We thus find that dS solutions, if they exist at all in this class of models, are perturbatively unstable. In fact, we could have already seen this directly from \eqref{pot2}. Indeed, the potential only has two differently scaling terms with respect to the $\beta$ modulus, both with negative exponents: $V=A \beta^{-a}+B \beta^{-b}$ with $a,b>0$. It is easy to check that a dS minimum would require a potential with at least \emph{three} differently scaling terms.
We thus conclude that at best AdS and Minkowski vacua are possible.

We will not study such vacua in detail here but only make a few remarks.
In the simplest case $F_0=T=0$, which applies to the smeared CDT1 model, one verifies that the potential only has runaway solutions unless $\hat{\mathcal{R}}_{10-d}=0$ and $V=0$ off-shell. Hence, only Minkowski vacua are possible in this case at the 4-derivative level. A preliminary analysis of more general setups with non-zero $F_0$ and/or $T$ furthermore suggests that self-consistent regimes (i.e., fine-tuned $C$, small $g_s$, large volume, small backreaction) are difficult to obtain. Finally, recall that there is in addition the issue of an open-string tachyon which may destabilize putative Minkowski or AdS vacua.

One may wonder whether these various issues can be overcome by including further corrections to the O8/D8 actions beyond the 4-derivative level or higher-derivative corrections to the bulk action.
We will not study such a scenario here since our main motivation was to consider a \emph{minimal} extension of the model of \cite{Cordova:2018dbb} which preserves its appealing simplicity and at the same time resolves the issues pointed out in \cite{Cribiori:2019clo}. Unfortunately, we found that this is impossible within the setting we considered. On the other hand, in a scenario where further corrections at even higher orders become relevant, this simplicity would clearly be lost. Moreover, one would then require even more fine-tuning in order to balance the various terms from different orders in the $\alpha^\prime$ expansion.

One might also object that our analysis was based on the scalar potential in the smeared approximation so that backreaction effects might change the result. In order to avoid the dS no-go, backreaction effects would have to lead to a violation of \eqref{o8nogo3}, i.e., they would have to significantly change the vacuum energy and/or the mass of the $\beta$ modulus. However, recall that the 10D no-go argument reviewed at the beginning of Section \ref{sec:o8nogo} holds without making assumptions on the warp factor, the dilaton or the internal metric and thus does not require a smeared limit. This implies that backreaction corrections do not affect
the vacuum energy classically. As explained before, the backreaction may create holes around the O8-planes in which string corrections are important (cf.~Fig.~\ref{smeared2}).
In regimes where these holes vanish and the curvatures/energy densities are small everywhere on the spacetime, the leading contribution to the vacuum energy should again be given by the O8/D8 4-derivative terms as in the smeared case. For finite but very small hole diameters, we expect this to still hold approximately, as discussed in Section \ref{sec:o8nogo}.
On the other hand, it is not obvious how the $\beta$ mass is affected by backreaction corrections.
To see this, one would have to study fluctuations in a fully backreacted 10D ansatz and derive the corresponding warped effective field theory, which is a quite involved task. We therefore leave a more detailed analysis of this question for future work \cite{Junghans:2024}.

We finally stress again that our results do not rule out dS vacua in a regime where singular holes cover a large part of the 10D spacetime.
In such a regime, the supergravity calculations performed here and in the previous works \cite{Cordova:2018dbb, Cribiori:2019clo, Cordova:2019cvf} are not applicable and a worldsheet analysis or other non-perturbative techniques would instead be required.

\section{Models with $p_i=6$ and $p_i=8$ sources}
\label{sec:o6}

In this section, we study models with $p_i=6$ and $p_i=8$ sources, i.e., O$6^\pm$-planes/D6-branes and O$8^\pm$-planes/D8-branes. Our main interest is in the CDT2 model, which was proposed in \cite{Cordova:2019cvf} and argued there to have dS solutions.
Our goal in this section is to repeat the analysis of Section \ref{sec:o8} for the CDT2 model and variants thereof with a similar ansatz. In particular, we again ask whether classical dS vacua are possible in these models and to what extent the result changes when we include $\alpha^\prime$ corrections to the O-plane/D-brane actions up to the 4-derivative order.

In Section \ref{sec:o6cdt2}, we will argue that the CDT2 model does not have dS or other vacua in regimes where the O-plane backreaction is small on most of the 10D spacetime so that the smeared approximation is valid. This holds both for classical and $\alpha^\prime$-corrected O-plane sources. In addition, we will present a second no-go argument against dS which does not make any assumptions about smearing and takes into account the full O-plane backreaction (analogously to the no-go for $p_i=8$ models reviewed in Section \ref{sec:o8nogo}). We will furthermore argue that the numerical dS solutions found in \cite{Cordova:2019cvf} avoid both no-go arguments by having non-standard source terms which are neither compatible with the classical nor the $\alpha^\prime$-corrected O-plane actions arising in string theory.

In Section \ref{sec:o6gen}, we will study generalized ``CDT2-like'' models in which we attempt to modify the assumptions on the geometry and the source and flux content such that the problems of the original model are avoided. However, we will see that, in the smeared approximation, the 4D scalar potential of such generalized models can at most have AdS or Minkowski vacua or unstable dS critical points but no dS vacua. This again holds both for classical source terms and including the 4-derivative corrections.

\subsection{The CDT2 model}
\label{sec:o6cdt2}

\subsubsection{Ansatz}
\label{sec:o6ansatz}

We start by reviewing the ansatz of \cite{Cordova:2019cvf}. The compactification space is a $\kappa_3 \times S^2$ fibration over an interval parametrized by $z$ with the metric
\begin{equation}
\d s_{10}^2= \e^{2A(z)} g_{\mu\nu}\d x^\mu \d x^\nu + \e^{-2A(z)} \left(\e^{2\lambda_3(z)} \d s_{\kappa_3}^2 + \e^{2\lambda_2(z)} \d s_{S^2}^2 + R^2 \d z^2 \right), \label{mt}
\end{equation}
where $\kappa_3$ is a negatively curved 3D Einstein space, $S^2$ is the unit 2-sphere, $R$ is a length scale and we take $z\in[0,\pi]$ without loss of generality.\footnote{
Note that $\e^{2A(z)}=\e^{2W(z)}$, $R^2=\e^{2q_0}$ and $z$ has a different range in the notation and conventions of \cite{Cordova:2019cvf}.}
The non-zero field strengths in the model are
\begin{equation}
F_0(z), \qquad F_2 = f_2(z) \text{dvol}_{S^2}, \qquad H_3 = h(z) \d z \w \text{dvol}_{S^2}. \label{fluxansatz}
\end{equation}
The localized sources are an O$6^-$-plane and a parallel anti-O$6^-$-plane wrapping $\kappa_3$ and an O$8^+$-plane wrapping $\kappa_3 \times S^2$, respectively.\footnote{Here we refer to the covering space. On the orientifold, the two O6-planes are identified and the O8-plane wraps $\kappa_3 \times \mathbb{RP}^2$.}
We remind the reader that ``anti'' refers to flipping the sign of the RR charge, while the $\pm$ superscripts refer to different signs of the tension (see \eqref{tension}).

The orientifold projection imposed in \cite{Cordova:2019cvf} mods out by $\Omega \sigma_8$ and $\Omega (-1)^{F_L}\sigma_6$, where $\Omega$ is the worldsheet parity operator, $(-1)^{F_L}$ is the left-moving fermion number and $\sigma_8$ and $\sigma_6$ are spacetime involutions defined as
\begin{equation}
\sigma_8: z \to \pi - z, \qquad \sigma_6: (\theta,\phi) \to (\pi-\theta,\phi+\pi),
\end{equation}
where $\theta$ and $\phi$ are the standard spherical coordinates on the $S^2$. Note that $\sigma_6$ is the antipodal map on the $S^2$, which implies that it does not have fixed points except at values of $z$ where the $S^2$ shrinks to a point. The fixed loci of $\sigma_6$ thus have codimension 3 and yield the O6-planes, while the fixed locus of $\sigma_8$ has codimension 1 and yields the O8-plane.

One verifies that $F_0\to -F_0$, $F_2\to F_2$, $H_3 \to -H_3$ under $\Omega$, $F_0\to -F_0$, $F_2\to -F_2$, $H_3 \to H_3$ under $(-1)^{F_L}$ \cite{Bergshoeff:2001pv} and $\d z \to - \d z$, $\text{dvol}_{S^2}\to -\text{dvol}_{S^2}$ under the spacetime involutions $\sigma_8$ and $\sigma_6$, respectively. We thus require
\begin{equation}
F_0(z) = -F_0(\pi-z), \qquad f_2(z) = f_2(\pi-z), \qquad h(z) = h(\pi-z) \label{s68}
\end{equation}
in order that $F_0$, $F_2$ and $H_3$ survive both projections $\Omega \sigma_8$ and $\Omega (-1)^{F_L}\sigma_6$.

\subsubsection{An obstruction to smeared vacua}
\label{sec:o6smear}

If we want to find dS vacua in a trustworthy regime where classical supergravity is meaningful, then a good place to start looking is a regime where the O-plane backreaction is small on most of the spacetime so that the solution is approximately smeared. According to our discussion in Section \ref{sec:smear}, the singular holes surrounding the O-planes are then very small and are thus expected to have a negligible effect on the 4D effective field theory and in particular the vacuum energy. However, we will now argue that the CDT2 model actually does not have any vacuum solutions in the smeared regime (except for the trivial solution where all fluxes, curvature and source terms are set to zero). Hence, backreaction effects of at least one of the O-planes are inevitably large on a large part of the spacetime. We will show that this is true both for classical source terms and including 4-derivative corrections.

Let us first discuss the metric \eqref{mt}. In the smeared solution, the warp factor satisfies $\e^{2A(z)}= 1$ and the internal curvature is constant. However, these requirements do by themselves not uniquely fix the metric but are consistent with \eqref{mt}, e.g., for $\e^{2\lambda_3(z)}=1$, $\e^{2\lambda_2(z)}=\text{const.}$ (yielding the space $\kappa_3 \times S^2 \times S^1$) or $\e^{2\lambda_3(z)}=1$, $\e^{2\lambda_2(z)}=R^2\sin^2(z)$ (yielding the space $\kappa_3 \times S^3$). Other possibilities would be spaces where $\kappa_3$ is non-trivially fibered over the interval. As a simple example, if $\kappa_3$ were a round $S^3$ instead of a negatively curved Einstein space, choosing $\e^{2\lambda_3(z)}=R^2\sin^2(z)$, $\e^{2\lambda_2(z)}=\text{const.}$ would yield a product of round spheres $S^4 \times S^2$, which again has constant curvature. It is therefore a priori not obvious which compactification space arises in the smeared limit.

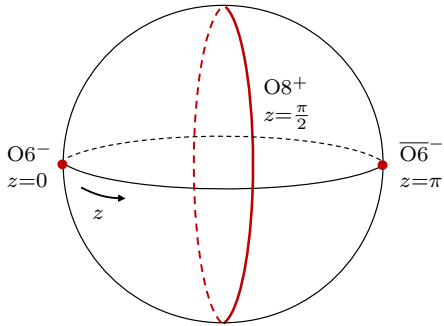
\begin{figure}[t!]
\centering
\includegraphics[trim = 0mm 50mm 0mm 40mm, clip, width=\textwidth]{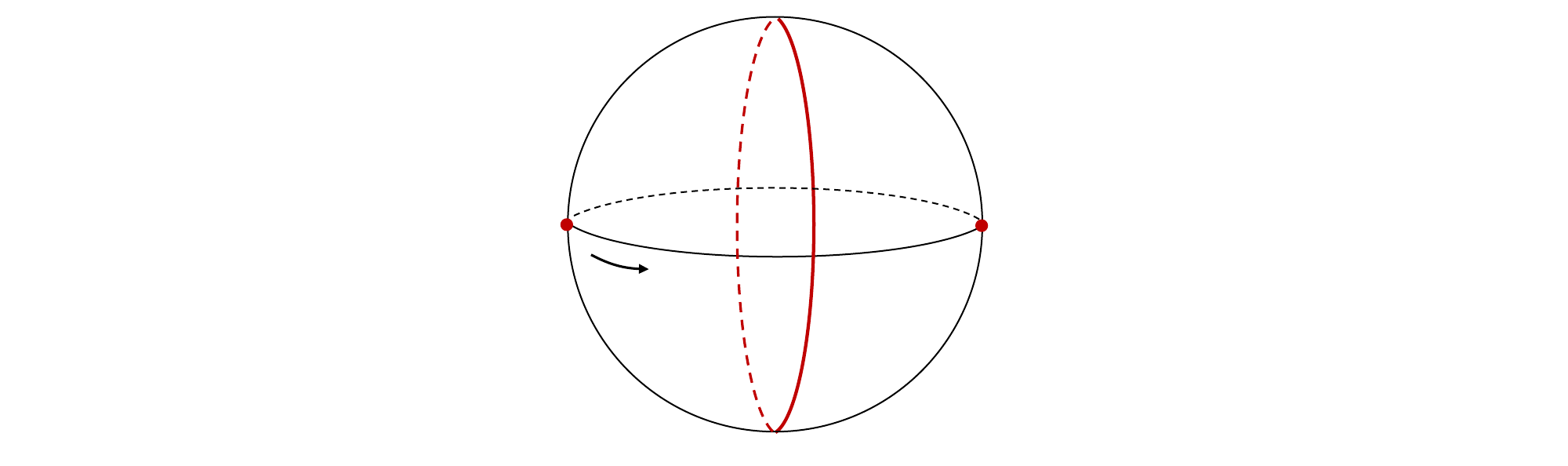}
\put(-310,71){$\scriptstyle{\text{O}6^-}$}
\put(-310,61){$\scriptstyle{z=0}$}
\put(-214,96){$\scriptstyle{\text{O}8^+}$}
\put(-214,86){$\scriptstyle{z=\frac{\pi}{2}}$}
\put(-163,71){$\scriptstyle{\overline{\text{O}6}^-}$}
\put(-163,61){$\scriptstyle{z=\pi}$}
\put(-278,49){$\scriptstyle{z}$}

\caption{$S^3$ with an O$6^-$-plane and an anti-O$6^-$-plane at $z=0$, $z=\pi$ (represented by the red dots) and an O$8^+$-plane wrapping an $S^2$ at $z=\frac{\pi}{2}$ (represented by the red circle).
\label{s3}}
\end{figure}

In the case at hand, a natural ansatz for the smeared metric is the second one, i.e., $\kappa_3 \times S^3$ with $\e^{2\lambda_3(z)}= 1$, $\e^{2\lambda_2(z)}= R^2\sin^2(z)$.
To see this, note that the backreaction of the O6-planes in \eqref{mt} makes the warp factor and the other functions depend on $z$ but not on the angles of the transverse $S^2$. This only makes sense if the $S^2$ shrinks to a point at the O6 positions in the unbackreacted space,
suggesting that the $S^2$ must remain non-trivially fibered over the interval and become an $S^3$ in the smeared solution, cf.~Fig.~\ref{s3}.
Indeed, as explained above, the O6-planes are the fixed points of the antipodal map on the $S^2$, which equal precisely the loci where the $S^2$ shrinks to a point. Hence, the presence of the O6-planes would be incompatible with constant finite $\e^{2\lambda_2}$ and we are naturally led to the assumption of a product space $\kappa_3 \times S^3$ in the smeared limit, where $z$ now corresponds to a polar angle in the usual spherical coordinates of the $S^3$.
Another way to arrive at the same conclusion is to note that the model has a single O8 as the only codimension-1 source. This would not be consistent with tadpole cancelation if the space were $\kappa_3 \times S^2 \times S^1$. However, no such problem arises if the O8 wraps a trivial cycle inside the $S^3$.
We thus conclude that the smeared limit of the metric \eqref{mt} is
\begin{equation}
\d s_{10}^2 = g_{\mu\nu}\d x^\mu \d x^\nu + \d s_{\kappa_3}^2 + R^2 \left(\sin^2(z)\, \d s_{S^2}^2 + \d z^2\right). \label{smearedmetric}
\end{equation}

Let us move on to the fluxes. In the smeared solution, no point on the $S^3$ is distinguished and therefore $F_0$, $|F_2|^2$ and $|H_3|^2$ must be constants. Using \eqref{fluxansatz}, \eqref{s68} and \eqref{smearedmetric}, this implies
\begin{equation}
F_0 = 0, \qquad F_2(z) = \tilde f_2 \sin^2(z) \d \text{vol}_{S^2}, \qquad H_3 = \tilde h \sin^2(z) \d z \w \d \text{vol}_{S^2} \label{fluxansatz2}
\end{equation}
for some constants $\tilde f_2$, $\tilde h$. The $F_2$ Bianchi identity (see \eqref{eom4}) becomes
\begin{equation}
\d F_2 =0 \label{f2bsmeared}
\end{equation}
in the smeared limit, where we used $F_0=0$ and the fact that the zero modes of the O6-plane and anti-O6-plane sources cancel out. Substituting \eqref{fluxansatz2} into \eqref{f2bsmeared} yields $\tilde f_2=0$ and thus the flux ansatz in the smeared limit is
\begin{equation}
F_0 = 0, \qquad F_2 = 0, \qquad H_3 = \tilde h \sin^2(z) \d z \w \d \text{vol}_{S^2}.
\end{equation}
We thus learn that $F_0$ and $F_2$ are purely generated by the O-plane backreaction in this model, whereas $H_3$ is a topological flux on the $S^3$.

We are now ready to study the equations of motion (see App.~\ref{app:eom0}). If the CDT2 model admits a regime where the O-plane backreaction becomes negligible,
then by assumption at most the zero modes of the O-plane sources can be relevant in the leading-order equations whereas the effect of all higher modes must be subleading (except at very small distances to the O-planes). This suggests that the limiting behavior is obtained by replacing
\begin{equation}
\delta(\Sigma_{\text{O}6^-}),\delta(\Sigma_{\overline{\text{O}6}^-}) \to \frac{1}{V_{S^3}}=\frac{1}{2\pi^2R^3}, \qquad \delta(\Sigma_{\text{O}8^+}) \to \frac{V_{S^2}}{V_{S^3}}=\frac{2}{\pi R} \label{repl}
\end{equation}
in the equations of motion, where $R$ is the $S^3$ radius, $V_{S^2}$ is the volume of the wrapped $S^2$ at the O8-plane locus $z=\frac{\pi}{2}$, and we used that $\int_{S^3} \d^3 y \sqrt{g_{S^3}}\, \delta(\Sigma_{\text{O}8^+})= V_{S^2}$.
Using \eqref{repl} and the above ansatz for the metric and the fluxes in \eqref{eom3}, the $zz$ and $S^2$ components of the Einstein equations become
\begin{align}
\R_{zz} &= \frac{g_{zz}}{8} \left(3|H_3|^2
- \frac{14 \e^\phi}{ \pi^2 R^3} + \frac{144 \e^\phi}{\pi R}\right), \label{eq:Einstein1} \\
\R_{ab} &= \frac{g_{ab}}{8} \left(3|H_3|^2
- \frac{14 \e^\phi}{\pi^2 R^3} + \frac{16 \e^\phi}{\pi R}\right), \label{eq:Einstein2}
\end{align}
where $a,b$ run over the two $S^2$ angles $\theta$ and $\phi$. Note that the $\frac{\e^\phi}{R^3}$ terms in the two equations are due to the tension of the two O6-planes and the $\frac{\e^\phi}{R}$ terms are due to the tension of the O8-plane.

As argued before, the $S^2$ is fibered over $z$ such that a round $S^3$ is obtained in limits of vanishing backreaction, which implies $\R_{ab} = \frac{2}{R^2}g_{ab}$ and $\R_{zz} = \frac{2}{R^2} g_{zz}$. However, this is incompatible with \eqref{eq:Einstein1}, \eqref{eq:Einstein2} unless the O8-plane terms $\sim\frac{\e^\phi}{R}$ become negligible compared to the leading terms in this limit.
Since for perturbative control we require $R\gg 1$, we furthermore have $\frac{\e^\phi}{R^3}\ll \frac{\e^\phi}{R}$ and therefore the O6-plane terms must also become negligible.
Hence, the limiting behavior of the O-plane sources in the CDT2 model is not \eqref{repl} but rather
\begin{equation}
\delta(\Sigma_{\text{O}6^-}),\delta(\Sigma_{\overline{\text{O}6}^-}), \delta(\Sigma_{\text{O}8^+}) \to 0. \label{prescription}
\end{equation}
We have thus shown that the CDT2 model does not admit a smeared limit where the O-plane sources are effectively replaced by their zero modes. Instead, the O-plane backreaction is inevitably large in this model unless \emph{all} Fourier modes of the O-plane sources -- including their zero modes -- become subleading in the equations of motion.

This conclusion is also supported by the $F_0$ Bianchi identity (see \eqref{eom4}), which using the prescription \eqref{prescription} becomes
\begin{equation}
\d F_0 = 0. \label{f0bianchi0}
\end{equation}
Note that this is consistent with our earlier result $F_0=0$. On the other hand, naively replacing the localized O8-plane term in \eqref{eom4} by the smeared one would give
\begin{equation}
\d F_0 \sim \d z, \label{f0bianchi}
\end{equation}
which yields a non-zero and varying $F_0$.
We thus observe that the zero mode of the O8-plane source term generates a backreaction correction to $F_0$ and should therefore not be included in the approximation where the backreaction is neglected. The $F_0$ Bianchi identity thus confirms that the correct prescription for the limit of small backreaction is indeed \eqref{prescription}, in agreement with what we concluded from the Einstein equations.

We now use this result in the dilaton and Einstein equations, which are stated in \eqref{eom1}--\eqref{eom3}. Substituting the ansatz for the metric and the fluxes and using \eqref{prescription}, we find
\begin{align}
0 &= - \mathcal{R}_4 - \frac{1}{2} |H_3|^2, \label{x1} \\
0 &= 2 \mathcal{ R}_4 + 2\mathcal{R}_{\kappa_3} + 2\mathcal{R}_{S^3} - |H_3|^2, \\
0 &= - \mathcal{R}_{\kappa_3} - \frac{3}{8}|H_3|^2, \label{x2} \\
0 &= - \mathcal{R}_{S^3} + \frac{9}{8} |H_3|^2.
\end{align}
Here we denote by $\mathcal{R}_4$, $\mathcal{R}_{\kappa_3}$ and $\mathcal{R}_{S^3}$ the scalar curvatures of the 4D external spacetime, the 3D Einstein space $\kappa_3$ and the 3-sphere with radius $R$, respectively.

It is straightforward to check that the only solution is the trivial one where all terms are set to zero:
\begin{equation}
\mathcal{R}_4 = \mathcal{R}_{\kappa_3}=\mathcal{R}_{S^3}=|H_3|^2=0.
\end{equation}
Hence, the CDT2 model does not have (non-trivial) vacuum solutions in regimes where backreaction effects are negligible.

Although we assumed classical source terms in the above arguments, it is clear that the same arguments apply if we turn on 4-derivative (or higher) couplings on the O-planes since, in the regime of small curvature/field strengths, these are subleading to the classical source terms. In particular, since we found that the
classical source terms $\frac{\e^\phi}{R}$ and $\frac{\e^\phi}{R^3}$ must be negligible in \eqref{eq:Einstein1}, \eqref{eq:Einstein2}, so are any 4-derivative corrections to them, and thus nothing changes in our analysis.
One might object that the $\alpha^\prime$ expansion breaks down close to the O-planes so that we should not assume that localized 4-derivative terms are subleading compared to the classical source terms. However, recall that we are in the smeared limit where the singular holes vanish and only the \emph{zero modes} of the O-plane sources appear in \eqref{eq:Einstein1}, \eqref{eq:Einstein2}, which have support over the whole 10D spacetime. Allowing the zero modes of the 4-derivative corrections to be larger than the zero modes of the classical source terms would therefore imply that the $\alpha^\prime$ expansion breaks down \emph{everywhere} in the 10D spacetime, not just very close to the O-planes, which would certainly not be a regime where we can trust supergravity. We therefore conclude that, even with $\alpha^\prime$-corrected source terms, there are no reliable dS vacua in the CDT2 model in the smeared regime.

Nevertheless, the reader may be concerned that our results do not fully rule out classical and almost classical dS in the CDT2 model.
In particular, we have not presented rigorous proofs for all of our claims (such as the behavior of the metric in the smeared limit).
Furthermore, as explained in Section \ref{sec:smear}, the existence of a regime where the solution is approximately smeared is useful but may not always be required to avoid large singular holes and a loss of control over the classical supergravity approach. In particular, we do not expect an issue with singularities for the O$8^+$-plane in the CDT2 model since it has a positive tension and should therefore not generate a hole. One might therefore imagine a situation where the backreaction of the two O$6^-$-planes is small on most of the spacetime so as to avoid large singular holes but the O$8^+$ backreaction is still important without creating a control problem.
Such solutions are not ruled out by the arguments in this section.
In the following, we will therefore discuss a second no-go argument for the CDT2 model, which does not rely on any assumptions about smearing and holds including the full non-linear backreaction of the O6/O8-planes. We will see that this alternative argument is in full agreement with our above conclusions and in particular again forbids classical dS vacua as well as dS vacua with $\alpha^\prime$-corrected source terms in regimes where supergravity can be trusted on most of the spacetime.

\subsubsection{Classical no-go}
\label{sec:o6class}

We now derive a no-go theorem against classical dS solutions in the CDT2 model which takes into account the full backreaction of the O6/O8-planes. Our strategy is similar to the no-go theorem of \cite{Cribiori:2019clo} for the O8/D8 models, which we reviewed in Section \ref{sec:o8nogo}. The idea is again to find a convenient combination of the 10D equations of motion (stated in App.~\ref{app:eom0}) which when integrated over the compact space yields a simple expression for the vacuum energy. We will assume the classical equations of motion and classical source terms in this section. As we will explain further below, this assumption is consistent as long as the singular holes in which classical supergravity breaks down are small.

We start by combining the dilaton and Einstein equations such that the 4D curvature is related to the curvatures of either $\kappa_3$ or the $S^2$ and a derivative term. In particular, one verifies using \eqref{eom1}--\eqref{eom3} that
\begin{align}
\e^{-2A} \mathcal{R}_4 &= \frac{4}{3} \e^{2A-2\lambda_3}\mathcal{R}_{\kappa_3} + 4 \frac{\e^{-4A+2\phi}}{\sqrt{g_6}} \partial_m\left( \e^{4A-2\phi} \sqrt{g_6}\, \partial^m\left(2A-\lambda_3\right)\right) \nll
= - \e^{2A-2\lambda_2} + \frac{\e^{-4A+2\phi}}{\sqrt{g_6}} \partial_m\left( \e^{4A-2\phi} \sqrt{g_6}\, \partial^m\left(3A+\lambda_2-\phi\right)\right) \label{dkddg}
\end{align}
holds for every solution satisfying the ansatz of Section \ref{sec:o6ansatz}. The first term in the second line is due to the curvature of the $S^2$.
Note that the classical O6/O8 source terms cancel out in both equations. To derive these equations, we used that the O6 sources wrap $\kappa_3$, that the O8 sources wrap $\kappa_3$ and the $S^2$ and that $F_2\sim \text{dvol}_{S^2}$, $H_3\sim \d z \w \text{dvol}_{S^2}$ by assumption of the ansatz, and we evaluated the various curvatures and covariant derivatives in \eqref{eom1}--\eqref{eom3} using the metric \eqref{mt}. The metric determinant $g_6$ is taken with respect to the six internal components of \eqref{mt}.

Integrating \eqref{dkddg} over the internal space, we obtain
\begin{align}
\left(\int\d^6 y \sqrt{g_6}\,  \e^{2A-2\phi} \right) \mathcal{R}_4 &=
\frac{4}{3} \left(\int\d^6 y \sqrt{g_6}\, \e^{6A-2\lambda_3-2\phi} \right) \mathcal{R}_{\kappa_3} =
- \int\d^6 y \sqrt{g_6}\, \e^{6A-2\lambda_2-2\phi} \label{ov2}
\end{align}
or, using \eqref{mt},
\begin{align}
\left(\int\d z\,  \e^{-4A+3\lambda_3+2\lambda_2-2\phi} \right) \mathcal{R}_4 &= \frac{4}{3} \left(\int\d z\, \e^{\lambda_3+2\lambda_2-2\phi} \right) \mathcal{R}_{\kappa_3} = - \int\d z\, \e^{3\lambda_3-2\phi}, \label{ov2b}
\end{align}
which relates the vacuum energy to the scalar curvatures of $\kappa_3$ and the $S^2$. Since the dS solutions in the CDT2 model of \cite{Cordova:2019cvf} have ${\mathcal{R}}_{\kappa_3}<0$, we observe that they are in conflict with both equations in \eqref{ov2b}. We instead find that only AdS solutions are allowed classically in the CDT2 model. An immediate question that comes to mind is whether dS is possible in alternative models where ${\mathcal{R}}_{\kappa_3}>0$ and the $S^2$ is replaced by some $M_2$ with ${\mathcal{R}}_{M_2}<0$. We will come back to such ``CDT2-like'' models in Section \ref{sec:o6gen}. As a cross-check of our result, note that the first part of \eqref{ov2b} becomes $\mathcal{R}_4 = \frac{4}{3} \mathcal{R}_{\kappa_3}$ in the smeared limit, which agrees with what one obtains when combining the smeared equations of motion \eqref{x1} and \eqref{x2} such that the $|H_3|^2$ term cancels out.

One might object that the above equations do not make sense since we integrated classical equations over the whole internal space including the singular holes surrounding the O6-planes where classical supergravity breaks down. Indeed, such a criticism was emphasized in \cite{Cordova:2019cvf} regarding the no-go theorem of \cite{Cribiori:2019clo}, which is based on a very similar integral expression (see Section \ref{sec:o8nogo}).

Let us therefore explain our philosophy in more detail. A useful way to think about the above argument is to only integrate \eqref{dkddg} up to a boundary, which is chosen such that it cuts out the singular holes and classical supergravity is reliable at every point we integrate over.
For example, we can define the boundary such that it excises all points in which curvatures and energy densities are larger than $0.1$ in string units (the precise number does not matter for the argument). This removes the two near-O6 regions from the integral and replaces them with two boundary terms in each equation in \eqref{ov2b}, which are functions of the hole diameter $\epsilon$. Since the solution at $z>\frac{\pi}{2}$ is just a copy of the solution at $z<\frac{\pi}{2}$ due to the orientifold involution, the two boundary terms are equal in each equation so that we can write them as a single boundary term for simplicity. Note that the O$8^+$ does not yield a further hole that needs to be cut out since it has a positive tension (cf.~the discussion in Section \ref{sec:smear}). We thus obtain
\begin{align}
\left(\int\limits_\epsilon^{\pi-\epsilon}\d z\,  \e^{-4A+3\lambda_3+2\lambda_2-2\phi} \right) \mathcal{R}_4 &=
\frac{4}{3} \left(\int\limits_\epsilon^{\pi-\epsilon}\d z\, \e^{\lambda_3+2\lambda_2-2\phi} \right) \mathcal{R}_{\kappa_3} + \mathcal{B}^{(1)}(\epsilon)
\nll = - \int\limits_\epsilon^{\pi-\epsilon}\d z\, \e^{3\lambda_3-2\phi} + \mathcal{B}^{(2)}(\epsilon), \label{ov2c}
\end{align}
where the coordinate $z$ is defined such that the ``centers'' of the holes are at $z=0$ and $z=\pi$.

Now we consider a $g_s\to 0$ limit as described in Section \ref{sec:smear} such that the singular holes and, consequently, our boundary shrink to points ($\epsilon\to 0$) and classical supergravity is reliable everywhere. We stay agnostic about the details of the limit (i.e., whether other fields aside from the dilaton are taken to scale non-trivially in the limit) and only demand that it is a limit where the O6 backreaction becomes small so that the holes vanish. Since we have already seen in Section \ref{sec:o6smear} that the backreaction of the O$8^+$ cannot be made small in the CDT2 model, we allow it to remain finite, i.e., the warp factor and the other functions can vary significantly over $z$. Note that, as discussed in Section \ref{sec:smear}, an O8 with a large transverse volume can backreact more strongly than an O6 so that this  is a priori not excluded.

Approaching the described limit, the boundary terms must become negligible in \eqref{ov2c}, i.e., $\lim_{\epsilon\to 0}\frac{\mathcal{B}^{(i)}}{\left(\int \d z\,  \e^{-4A+3\lambda_3+2\lambda_2-2\phi} \right) \mathcal{R}_4} = 0$. Hence, for sufficiently small $\epsilon$ and restricting to classical sources, the holes do not affect the classical no-go. One might wonder whether the O6-planes that sit in the regions excised by the boundary could create some discontinuity such that the boundary terms do not vanish in the limit. However, this is just another way of saying that the O6-planes contribute source terms to \eqref{dkddg}, and in fact these combinations of the equations of motion are precisely such that no such sources appear at the classical level. The leading contribution of the O6/O8-planes to \eqref{ov2} thus arises earliest at the 4-derivative order. As will be discussed in the next subsection, these corrections are generically non-vanishing but too small to circumvent the no-go in the small-hole regime.

On the other hand, if we take $g_s$ large or the volume small so that the holes eat up a large part of the naive spacetime (i.e., the Small-Hole Condition of Section \ref{sec:smear} is violated as on the left-hand side of Fig.~\ref{smeared}), string corrections to \eqref{ov2} can be important and we can no longer make a reliable statement about the possibility of dS. However, in that case, we cannot trust classical supergravity in the first place, and in particular using the classical equations of motion to look for dS vacua as in \cite{Cordova:2019cvf} is not a meaningful approach anymore.
In such a regime, one should instead use worldsheet or other non-perturbative methods to compute the vacuum energy. We thus conclude that \eqref{ov2} rules out dS vacua in the CDT2 model in regimes where an approach based on classical supergravity is meaningful.

\subsubsection{Adding $\alpha^\prime$ corrections}
\label{sec:o6alpha}

Let us now include the 4-derivative corrections \eqref{la} to the O-plane source terms in \eqref{dkddg}. This yields
\begin{align}
\e^{-2A} \mathcal{R}_4 &= \frac{4}{3} \e^{2A-2\lambda_3}\mathcal{R}_{\kappa_3} + 4 \frac{\e^{-4A+2\phi}}{\sqrt{g_6}} \partial_m\left( \e^{4A-2\phi} \sqrt{g_6}\, \partial^m\left(2A-\lambda_3\right)\right) \nl - \sum_i \frac{ T_i }{2\pi} \e^{2\phi}\delta(\Sigma_i) \left(\frac{4}{3}\frac{\delta(\e^{-\phi}\mathcal{L}_{\alpha^{\prime 2},i})}{\delta g^{xy}} g^{xy} - \frac{\delta(\e^{-\phi}\mathcal{L}_{\alpha^{\prime 2},i})}{\delta g^{\mu\nu}} g^{\mu\nu} \right) \nll
= - \e^{2A-2\lambda_2} + \frac{\e^{-4A+2\phi}}{\sqrt{g_6}} \partial_m\left( \e^{4A-2\phi} \sqrt{g_6}\, \partial^m\left(3A+\lambda_2-\phi\right)\right) \nl - \sum_i \frac{ T_i }{2\pi} \e^{2\phi}\delta(\Sigma_i) \left( \frac{1}{4} \e^{-\phi}\mathcal{L}_{\alpha^{\prime 2},i} +
\frac{1}{4} \frac{\delta (\e^{-\phi}\mathcal{L}_{\alpha^{\prime 2},i})}{\delta \phi}
+\frac{1}{2}\frac{\delta (\e^{-\phi}\mathcal{L}_{\alpha^{\prime 2},i})}{\delta g^{xy}} g^{xy} \right. \nl \left.
+\frac{1}{2} \frac{\delta (\e^{-\phi}\mathcal{L}_{\alpha^{\prime 2},i})}{\delta g^{zz}} g^{zz} 
- \frac{1}{2}\frac{\delta (\e^{-\phi}\mathcal{L}_{\alpha^{\prime 2},i})}{\delta g^{\mu\nu}} g^{\mu\nu} \right),
\end{align}
where we denote by $x,y$ the internal indices parallel to $\kappa_3$.
We will also need a third combination of the equations of motion,
\begin{align}
\e^{-2A} \mathcal{R}_4 &= - \e^{2\phi}F_0^2 - \e^{4A-4\lambda_2+2\phi}f_2^2  +4\frac{\e^{-4A+2\phi}}{\sqrt{g_6}} \partial_m\left( \e^{4A-2\phi} \sqrt{g_6}\, \partial^m A \right) \nl - \sum_i \frac{ T_i }{2\pi} \e^{2\phi}\delta(\Sigma_i) \left(\e^{-\phi}+ 2\e^{-\phi}\mathcal{L}_{\alpha^{\prime 2},i} + \frac{\delta (\e^{-\phi}\mathcal{L}_{\alpha^{\prime 2},i})}{\delta \phi}
- \frac{\delta (\e^{-\phi}\mathcal{L}_{\alpha^{\prime 2},i})}{\delta g^{\mu\nu}} g^{\mu\nu} \right).
\end{align}
Note that the sources in the first two equations appear earliest at the 4-derivative level, whereas the third equation also has classical sources.
As in \cite{Cordova:2019cvf}, we assume that $\alpha^\prime$ corrections in the bulk spacetime are negligible.

We now integrate the above equations over the internal space as before (with weight $\e^{4A-2\phi}\sqrt{g_6}$), where we again choose a 
boundary such that the unreliable holes near the O6-planes are cut out. Using \eqref{mt}, this yields
\begin{align}
\bigg(\int & \d z\, \e^{-4A+3\lambda_3+2\lambda_2-2\phi} \bigg) \mathcal{R}_4 \nll =
\frac{4}{3} \left(\int \d z\, \e^{\lambda_3+2\lambda_2-2\phi} \right) \mathcal{R}_{\kappa_3} + \mathcal{B}^{(1)}(\epsilon) \nl
- \frac{ T_{\text{O}8^+} }{2\pi R} \left.\e^{-A+3\lambda_3+2\lambda_2} \left(\frac{4}{3}\frac{\delta (\e^{-\phi}\mathcal{L}_{\alpha^{\prime 2},{\text{O}8^+}})}{\delta g^{xy}} g^{xy} - \frac{\delta (\e^{-\phi}\mathcal{L}_{\alpha^{\prime 2},{\text{O}8^+}})}{\delta g^{\mu\nu}} g^{\mu\nu} \right) \right|_{z=\frac{\pi}{2}}  \label{ov3b}
 \\
&= - \int\d z\, \e^{3\lambda_3-2\phi} + \mathcal{B}^{(2)}(\epsilon) \nl - \frac{ T_{\text{O}8^+} }{2\pi R} \left.\e^{-A+3\lambda_3+2\lambda_2} \left(
\frac{1}{4} \e^{-\phi}\mathcal{L}_{\alpha^{\prime 2},{\text{O}8^+}}
+ \frac{1}{4} \frac{\delta (\e^{-\phi}\mathcal{L}_{\alpha^{\prime 2},{\text{O}8^+}})}{\delta \phi} 
+\frac{1}{2}\frac{\delta (\e^{-\phi}\mathcal{L}_{\alpha^{\prime 2},{\text{O}8^+}})}{\delta g^{xy}} g^{xy} \right.\right. \nl \left.\left.
+\frac{1}{2} \frac{\delta (\e^{-\phi}\mathcal{L}_{\alpha^{\prime 2},{\text{O}8^+}})}{\delta g^{zz}} g^{zz} 
- \frac{1}{2}\frac{\delta (\e^{-\phi}\mathcal{L}_{\alpha^{\prime 2},{\text{O}8^+}})}{\delta g^{\mu\nu}} g^{\mu\nu} \right) \right|_{z=\frac{\pi}{2}} \label{ov3c} \\
&= - \left(\int \d z\, \e^{-2A+3\lambda_3+2\lambda_2} \right) F_0^2 - \int\d z\, \e^{2A+3\lambda_3-2\lambda_2} f_2^2  + \mathcal{B}^{(3)}(\epsilon) \nl
- \frac{ T_{\text{O}8^+} }{2\pi R} \left.\e^{-A+3\lambda_3+2\lambda_2} \left(\e^{-\phi}+ 2\e^{-\phi}\mathcal{L}_{\alpha^{\prime 2},{\text{O}8^+}} + \frac{\delta (\e^{-\phi}\mathcal{L}_{\alpha^{\prime 2},{\text{O}8^+}})}{\delta \phi} - \frac{\delta (\e^{-\phi}\mathcal{L}_{\alpha^{\prime 2},{\text{O}8^+}})}{\delta g^{\mu\nu}} g^{\mu\nu} \right) \right|_{z=\frac{\pi}{2}}, \label{ov3}
\end{align}
where the integrals are taken in the interval $z\in[\epsilon,\pi-\epsilon]$ and the boundary terms are given by
\begin{align}
\mathcal{B}^{(1)}(\epsilon) &= -\frac{8}{R^2} \left. \e^{3\lambda_3+2\lambda_2-2\phi} \left(2A-\lambda_3\right)^\prime\right|_{z=\epsilon}, \\
\mathcal{B}^{(2)}(\epsilon) &= - \frac{2}{R^2}\left. \e^{3\lambda_3+2\lambda_2-2\phi} \left(3A+\lambda_2-\phi\right)^\prime\right|_{z=\epsilon}, \\
\mathcal{B}^{(3)}(\epsilon) &= -\frac{8}{R^2} \left. \e^{3\lambda_3+2\lambda_2-2\phi} A^\prime \right|_{z=\epsilon}
\end{align}
with $'=\frac{\d}{\d z}$. We thus see that, if the O8 and boundary terms in \eqref{ov3b} and \eqref{ov3c} are positive and large enough, the classical dS no-go may in principle be avoided. To see whether this is possible, we have to impose boundary conditions near the O6s.

By Gauss's law, the boundary terms equal integrals over whatever is inside the cut-out holes but since these regions are expected to be non-perturbative, it is a priori not obvious how to evaluate them. However, this does not mean that anything goes. In particular, in the $g_s\to 0$ limit where the O6 backreaction vanishes and the holes shrink to points, the contribution of each boundary integral is expected to equal that of a probe O6-plane. The contribution of such an O6-plane is not arbitrary but fixed by string scattering amplitudes and T-duality arguments \cite{Bachas:1999um, Wyllard:2000qe, Wyllard:2001ye, Fotopoulos:2001pt, Schnitzer:2002rt, Garousi:2006zh, Garousi:2009dj, Robbins:2014ara, Garousi:2014oya}. We therefore claim that, for small holes, the correct boundary conditions are
\begin{align}
\mathcal{B}^{(1)}(\epsilon) &=
-2\frac{ T_{\text{O}6^-} }{8\pi^2 R} \e^{A+3\lambda_3} \left.\left(\frac{4}{3}\frac{\delta (\e^{-\phi}\mathcal{L}_{\alpha^{\prime 2},{\text{O}6^-}})}{\delta g^{xy}} g^{xy} - \frac{\delta (\e^{-\phi}\mathcal{L}_{\alpha^{\prime 2},{\text{O}6^-}})}{\delta g^{\mu\nu}} g^{\mu\nu} \right)\right|_{z=\epsilon}, \\
\mathcal{B}^{(2)}(\epsilon) &= -2\frac{ T_{\text{O}6^-} }{8\pi^2 R} \e^{A+3\lambda_3} \left.\left(
\frac{1}{4} \e^{-\phi}\mathcal{L}_{\alpha^{\prime 2},{\text{O}6^-}}
+ \frac{1}{4} \frac{\delta (\e^{-\phi}\mathcal{L}_{\alpha^{\prime 2},{\text{O}6^-}})}{\delta \phi}
+\frac{1}{2}\frac{\delta (\e^{-\phi}\mathcal{L}_{\alpha^{\prime 2},{\text{O}6^-}})}{\delta g^{xy}} g^{xy}\right.\right. \nl \left.\left.
+\frac{1}{2} \frac{\delta (\e^{-\phi}\mathcal{L}_{\alpha^{\prime 2},{\text{O}6^-}})}{\delta g^{zz}} g^{zz} 
- \frac{1}{2}\frac{\delta (\e^{-\phi}\mathcal{L}_{\alpha^{\prime 2},{\text{O}6^-}})}{\delta g^{\mu\nu}} g^{\mu\nu} \right)\right|_{z=\epsilon}, \\
\mathcal{B}^{(3)}(\epsilon) &= -2\frac{ T_{\text{O}6^-} }{8\pi^2 R} \e^{A+3\lambda_3} \left.\left( \e^{-\phi}+ 2\e^{-\phi}\mathcal{L}_{\alpha^{\prime 2},{\text{O}6^-}} + \frac{\delta (\e^{-\phi}\mathcal{L}_{\alpha^{\prime 2},{\text{O}6^-}})}{\delta \phi} - \frac{\delta (\e^{-\phi}\mathcal{L}_{\alpha^{\prime 2},{\text{O}6^-}})}{\delta g^{\mu\nu}} g^{\mu\nu} \right)\right|_{z=\epsilon}
\end{align}
up to higher-than-4-derivative corrections which are neglected here. In the limit $\epsilon\to 0$, this yields exactly the behavior predicted by string theory and should therefore still be approximately true at small finite $\epsilon$. On the other hand, we emphasize once more that one should not expect the above expressions to hold if $\epsilon=\mathcal{O}(1)$. Indeed, as $g_s$ is increased or the volume decreased so that the O6 holes become larger,
the perturbative string amplitudes determining the O-plane action become less reliable and at the same time the supergravity description of the solution in the bulk spacetime breaks down as well. Instead of a clear distinction between a weakly curved bulk and localized holes containing O-planes, the solution as a whole then becomes a strongly curved, stringy blob and control is lost (cf.~Fig.~\ref{smeared}).

We also note that, in contrast to what we imposed above, the boundary conditions in the numerical dS solution of \cite{Cordova:2019cvf} leave $(2A-\lambda_3)^\prime$ and $(3A+\lambda_2-\phi)^\prime$ \emph{unspecified} near the holes so that the hole contribution to \eqref{ov3b}, \eqref{ov3c} in that solution is allowed to differ from that of an actual O6-plane. This is very similar to the permissive boundary conditions in the CDT1 model which leave the sources unspecified in those combinations of the equations of motion where the classical O8 tension cancels out, as discussed in Section \ref{sec:bc}. We claim, however, that $(2A-\lambda_3)^\prime$ and $(3A+\lambda_2-\phi)^\prime$ do in fact have to satisfy boundary conditions and that these boundary conditions are fixed precisely by the higher-derivative terms stated above.

Imposing this, \eqref{ov3b}--\eqref{ov3} become
\begin{align}
\bigg(\int \d z\, \e^{-4A+3\lambda_3+2\lambda_2-2\phi} \bigg) \mathcal{R}_4 &=
\frac{4}{3} \left(\int \d z\, \e^{\lambda_3+2\lambda_2-2\phi} \right) \mathcal{R}_{\kappa_3}
+ 2 \frac{ T_{\text{O}6^-} }{8\pi^2 R} \left. \e^{A+3\lambda_3-\phi} \mathcal{O}(E^2) \right|_{z=\epsilon} \nl
+ \frac{ T_{\text{O}8^+} }{2\pi R} \left.\e^{-A+3\lambda_3+2\lambda_2-\phi} \mathcal{O}(E^2) \right|_{z=\frac{\pi}{2}} \label{ov4a} \\
&= - \int\d z\, \e^{3\lambda_3-2\phi} + 2 \frac{ T_{\text{O}6^-} }{8\pi^2 R} \left. \e^{A+3\lambda_3-\phi} \mathcal{O}(E^2) \right|_{z=\epsilon} \nl
+ \frac{ T_{\text{O}8^+} }{2\pi R} \left.\e^{-A+3\lambda_3+2\lambda_2-\phi} \mathcal{O}(E^2) \right|_{z=\frac{\pi}{2}} \label{ov4b} \\
&= - \left(\int \d z\, \e^{-2A+3\lambda_3+2\lambda_2} \right) F_0^2 - \int\d z\, \e^{2A+3\lambda_3-2\lambda_2} f_2^2 \nl
- 2 \frac{ T_{\text{O}6^-} }{8\pi^2 R} \left. \e^{A+3\lambda_3-\phi} \left(1+ \mathcal{O}(E^2)\right) \right|_{z=\epsilon} \nl
- \frac{ T_{\text{O}8^+} }{2\pi R} \left.\e^{-A+3\lambda_3+2\lambda_2-\phi} \left(1+ \mathcal{O}(E^2)\right) \right|_{z=\frac{\pi}{2}}, \label{ov4c}
\end{align}
where all integrals are over $z\in[\epsilon,\pi-\epsilon]$. Here and in what follows, we schematically denote by $E^2$ all 4-derivative corrections due to $\mathcal{L}_{\alpha^{\prime 2},i}$, indicating with the notation that they are quadratic in the string-frame energy densities/curvatures. The precise form and the numerical coefficients of these terms will not be relevant for our argument.\footnote{The only required input is that the coefficients are $\lesssim \mathcal{O}(1)$, which is true according to, e.g., \cite{Robbins:2014ara}.}

We are now ready to analyze whether the 4-derivative corrections $\sim E^2$ can affect the classical dS no-go. According to \eqref{ov4a} and \eqref{ov4b}, the only solution is AdS unless the $E^2$ terms are large enough to compete with the integrated $S^2$ and $\kappa_3$ curvatures. One may already suspect at this point that it will be difficult to cancel a classical term with an $\alpha^\prime$ correction and at the same time ensure small curvature and energy densities. However, since the $\alpha^\prime$ corrections live at the O-plane loci whereas the classical terms are integrated over the whole bulk, warping effects might provide large factors and thus allow a balance between terms of different orders in the $\alpha^\prime$ expansion. We therefore have to analyze this possibility more carefully.

To this end, we first note that \eqref{ov4c} implies for dS that
\begin{equation}
2 \frac{ |T_{\text{O}6^-}| }{8\pi^2 R} \left. \e^{A+3\lambda_3-\phi} \right|_{z=\epsilon} \gtrsim \frac{ T_{\text{O}8^+} }{2\pi R} \left.\e^{-A+3\lambda_3+2\lambda_2-\phi} \right|_{z=\frac{\pi}{2}}.
\end{equation}
Hence, if we can show that the O6 $\alpha^\prime$ corrections are negligibly small in \eqref{ov4b} for arbitrary $E^2\ll 1$, then the same is also true for the O8 $\alpha^\prime$ corrections for arbitrary $E^2\ll 1$. We therefore focus on the effect of the O6 corrections from now on.
In particular, according to \eqref{ov4b}, dS would require
\begin{equation}
\frac{2}{\pi R}\left.\e^{A+3\lambda_3-\phi} E^2 \right|_{z=\epsilon} \gtrsim \int\d z\, \e^{3\lambda_3-2\phi}, \label{dsgdsdseekp}
\end{equation}
where we used $T_{\text{O}6^-}=-8\pi$.

Since we defined the hole boundary at $z=\epsilon$ such that supergravity is valid everywhere in the interval of integration, we demand that the string-frame curvatures and energy densities are weak and the string coupling is small there. This implies in particular that
\begin{equation}
\frac{\e^{2A}}{R^2}(A^\prime)^2 \ll 1, \qquad \frac{\e^{2A}}{R^2} (\lambda_3^\prime)^2 \ll 1, \qquad \frac{\e^{2A}}{R^2} (\phi^\prime)^2 \ll 1, \qquad E \ll 1, \qquad \e^\phi \ll 1 \label{zrzrziuo}
\end{equation}
for all $z\in[\epsilon,\pi-\epsilon]$. Since the first three conditions may not be obvious, note that the first two terms enter curvature invariants like the Ricci scalar while the third one is the energy density of the dilaton so that these three terms need to be small for control. If these terms are large, they will furthermore generate other large energy densities through the equations of motion, which we do not want.

In order to meaningfully talk about an effective supergravity solution, we also require that the relevant length scales and in particular the interval along $z$ are large in string units, which implies that the maximum of $\e^{-A}R$ on $z\in[\epsilon,\pi-\epsilon]$ must be $\gg 1$. Since $ |(\frac{\e^{A}}{R})^\prime| \ll 1$, this implies the stronger condition that
\begin{equation}
\e^{-A}R \gg 1
\end{equation}
for all $z\in[\epsilon,\pi-\epsilon]$.

Using \eqref{zrzrziuo}, we conclude that
\begin{equation}
\frac{1}{R}|(\e^{A+3\lambda_3-2\phi})^\prime| \ll \e^{3\lambda_3-2\phi}
\end{equation}
for all $z\in[\epsilon,\pi-\epsilon]$. This in turn implies
\begin{align}
& \frac{1}{R}\left(\e^{A+3\lambda_3-2\phi}|_{z=\epsilon}-\e^{A+3\lambda_3-2\phi}|_{z=z_\text{min}}\right) =\frac{1}{R} \left|\int_{\epsilon}^{z_\text{min}} \d z (\e^{A+3\lambda_3-2\phi})^\prime \right| \nll \le \frac{1}{R}\int_{\epsilon}^{z_\text{min}} \d z \left|(\e^{A+3\lambda_3-2\phi})^\prime\right|  \le 
\frac{1}{R}\int_\epsilon^{\pi-\epsilon} \d z \left|(\e^{A+3\lambda_3-2\phi})^\prime\right| \ll \int_\epsilon^{\pi-\epsilon} \d z\, \e^{3\lambda_3-2\phi}, \label{eiotertur}
\end{align}
where by $z_\text{min}$ we mean the value of $z\in[\epsilon,\pi-\epsilon]$ for which $\e^{A+3\lambda_3-2\phi}$ is at its global minimum. Since $\frac{1}{R}\e^{A+3\lambda_3-2\phi}|_{z=z_\text{min}} \le \frac{1}{R(\pi-2\epsilon)}\int_\epsilon^{\pi-\epsilon} \d z\, \e^{A+3\lambda_3-2\phi} \ll \int_\epsilon^{\pi-\epsilon} \d z\, \e^{3\lambda_3-2\phi}$, we can ignore the $\e^{A+3\lambda_3-2\phi}|_{z=z_\text{min}}$ term in \eqref{eiotertur}.
We thus arrive at
\begin{equation}
\frac{1}{R}\e^{A+3\lambda_3-2\phi}|_{z=\epsilon} \ll \int_\epsilon^{\pi-\epsilon} \d z\, \e^{3\lambda_3-2\phi}. \label{dshsdhjghg}
\end{equation}
Combining \eqref{dsgdsdseekp} and \eqref{dshsdhjghg} then finally yields
\begin{equation}
\e^{\phi}E^2|_{z=\epsilon} \gg 1,
\end{equation}
which is clearly inconsistent with the condition \eqref{zrzrziuo}.

We have thus shown that it is impossible in the CDT2 model to balance the $\alpha^\prime$ corrections from the O6-planes/hole regions with the classical bulk terms in such a way that we obtain a dS solution in which the curvature/energy densities and the string coupling are small over most of the bulk spacetime.
The reason is that the $\alpha^\prime$ corrections to the vacuum energy are suppressed compared to the negative classical terms by the string coupling and a factor quadratic in the curvatures/energy densities. Even if each of these factors is just moderately small at the boundary, say $\e^\phi\sim E \sim \mathcal{O}(10^{-1})$, they together give a large suppression, and we have seen that warping effects cannot consistently be made large enough to compensate this suppression since large variations in the fields inevitably lead to large curvature and energy densities somewhere in the bulk. Remarkably, this conclusion did not require us to know any specifics about the coefficients and the signs of the 4-derivative corrections.

In conclusion, the O6/O8 contributions to \eqref{ov4a}, \eqref{ov4b} are negligible and therefore \eqref{ov2} approximately holds in the regime where singular holes are small and classical supergravity is reliable on most of the spacetime. Taking into account 4-derivative (or higher) corrections to the O-plane source terms does therefore not affect our dS no-go for the CDT2 model.

\subsubsection{Non-standard sources}
\label{sec:o6source}

In the previous subsections, we argued using two complementary approaches that the CDT2 model does not admit dS vacua in regimes where classical supergravity is trustworthy.
Nevertheless, \cite{Cordova:2019cvf} found a numerical dS solution in this model, which suggests that one of the assumptions that went into our no-go is violated there.
In particular, the following two assumptions were crucial for our arguments:
\begin{itemize}
\item The singular holes surrounding the O-planes are small enough that classical supergravity is meaningful on most of the spacetime (Small-Hole Condition, cf.~the discussion in Section \ref{sec:smear}).
\item The source terms in the equations of motion are classical or almost classical, i.e., they are determined by the classical or $\alpha^\prime$-corrected O-plane action derived from string theory (cf.~\eqref{action2} and the discussion in Section \ref{sec:bc}).
\end{itemize}
On the other hand, if one allows singular holes that cover a large part of the spacetime or source terms that disagree with the known O-plane actions derived from string theory, then one can formally avoid our dS no-go.
Our interpretation is that such solutions should be considered unphysical since, in the first case, using supergravity is not self-consistent, while, in the second case, it is unclear whether the sources correspond to any objects that exist in string theory.
Let us also stress that satisfying the above two assumptions is of course not a guarantee for the non-perturbative existence of a supergravity solution but should rather be understood as a useful sanity check (i.e., we expect them to be necessary but not sufficient to trust the solution).

The dS solution of \cite{Cordova:2019cvf} appears to circumvent our no-go by violating our second assumption, i.e., by having non-standard source terms.
Indeed, it was observed in \cite[footnote 8]{Cordova:2019cvf} that the boundary conditions of the fields near the O6-planes in their dS solution are quite unusual. Specifically, the warp factor $\e^{-4A}$ does not diverge and the other warp factors $\e^{\lambda_i}$ and the dilaton $\e^\phi$ go to zero with unusual power laws as one approaches the center of the hole of one of the O6s.

In App.~\ref{app:o6}, we reproduce this behavior analytically by computing the local solution in an expansion around the locus of the putative O6-plane (following a similar computation performed in \cite[App.~B]{Blaback:2011pn} for a different compactification). We find that several solutions are consistent with the observed behavior in \cite[footnote 8]{Cordova:2019cvf}.\footnote{We assume a power-law behavior of the fields near the O6 as suggested by \cite[footnote 8]{Cordova:2019cvf} and can therefore not exclude that there are further boundary conditions with, e.g., logarithmic scalings, which would be even more exotic.} Choosing the source at $z=0$ without loss of generality, the warp factors of the metric and the dilaton satisfy
\begin{equation}
\e^{-4A} \sim z^{-F + \frac{\sqrt{-15F^2+48FM-60M^2-24F+24M+12}}{3}}, \quad \e^{\lambda_3}\sim z^{\frac{1}{3}(2F-2M+1)}, \quad
\e^{\lambda_2}\sim z^{M}, \quad \e^\phi \sim z^{F} \label{st1}
\end{equation}
with two free parameters $M$, $F$. We refer to App.~\ref{app:o6} for more details and the solution for the other fields $f_2$ and $h$.
The allowed ranges of $M$ and $F$ depend on the behavior of $f_2$ and are $\text{max}\left(0,M-\frac{1}{2}\right) < F \le M-\frac{1}{2}+\frac{\sqrt{3-6M^2}}{2}$, $0< M < \frac{1}{\sqrt{2}}$ if $f_2\to 0$ in the center of the hole and $\text{max}\left(0,\frac{4}{5}(2M-1) + \frac{\sqrt{-9M^2 - 6M + 9}}{10}\right) < F \le M-\frac{1}{2}+\frac{\sqrt{3-6M^2}}{2}$, $0< M < \frac{1+\sqrt{6}}{5}$ if $f_2$ remains finite.

Note in particular the unusual irrational exponents in \eqref{st1}, which are reminiscent of similar boundary conditions found in \cite[App.~B]{Blaback:2011pn}, whereas the standard flat-space O6 solution satisfies $\e^{-4A} \sim z^{-1}$, $\e^\phi \sim z^{\frac{3}{4}}$, $\e^{\lambda_3}\sim z^{\frac{1}{2}}$, $\e^{\lambda_2}\sim z$ in the hole.
The latter reference also derived the corresponding delta-function sources that would have to appear in the equations of motion in order to generate such boundary conditions with irrational exponents. The required source terms were found in \cite{Blaback:2011pn} to be such that they could not be associated to any known brane in string theory and were therefore classified as unphysical. It would be interesting to perform an analogous study for the boundary conditions given in \eqref{st1} and check that the corresponding source terms are not associated to an O6-plane, as predicted by our no-go arguments. We leave this for future work.

\subsection{Generalized models}
\label{sec:o6gen}

In Section \ref{sec:o6cdt2}, we argued from several different perspectives that the CDT2 model does not have dS vacua in the regime where a classical calculation is trustworthy.
We also argued that this cannot be remedied by turning on 4-derivative (or higher) corrections to the classical O-plane source terms. In what follows, we will attempt to improve this situation by studying generalizations of the CDT2 model which avoid some of the issues of the original model.
We will focus on the smeared approximation in this section for simplicity. The reason is that, similar to Section \ref{sec:o8nogo2}, our arguments are based on identifying an instability of the putative dS solutions, which would be technically very involved for the corresponding fully backreacted ansatz.
As explained before, looking for dS vacua in an approximately smeared regime is motivated by the desire to avoid large singular holes for the O$6^-$-planes but may not strictly be necessary for all types of sources (in particular, O$8^-$-planes and any type of positive-tension source).

There are several straightforward ways to modify the setup of the CDT2 model without completely losing its appealing simplicity. One possibility is to replace the $S^3$ transverse to the O6-planes by a different manifold $M_3$ such as the product manifold $M_3 = M_2 \times S^1$, where $M_2$ is a closed 2D manifold. In contrast to the (round) $S^3$ case, the Ricci curvature is in general not proportional to the metric for such a manifold so that the no-go argument of Section \ref{sec:o6smear} is circumvented.
In view of the no-go \eqref{ov2}, we will also drop the assumption that $\kappa_3$ has negative curvature and instead allow it to be a general closed 3D Einstein manifold.
In addition, we can turn on $F_0$ or $F_2$ flux on $M_2$, which gives us more freedom compared to the CDT2 model, which only has $H_3$ flux in the smeared limit. Note that it depends on the chosen orientifold involution which of these fluxes can consistently be turned on but we will keep the expressions general in the following. Finally, we allow an arbitrary number of O6-planes and/or D6-branes wrapping $\kappa_3$ and an arbitrary number of O8-planes/D8-branes wrapping $\kappa_3 \times M_2$.

For such ``CDT2-like'' models, we can compute the 4D scalar potential in the smeared approximation and ask whether dS vacua are possible, either with classical sources or including the 4-derivative corrections. As we will see below, this leads to the conclusion that any dS extrema that may exist in such models always have a tachyon.

\subsubsection{Scalar potential}
\label{sec:o6scal}

We now again compute the scalar potential as we did in Section \ref{sec:scal} for the O8/D8 models.
Our ansatz for the metric in the smeared limit is
\begin{equation}
\d s_{10}^2= g_{\mu\nu}\d x^\mu \d x^\nu + \d s_{\kappa_3}^2 + \d s_{M_3}^2. \label{dfjogrdj}
\end{equation}
We allow $\kappa_3$ to be a 3D Einstein manifold with arbitrary (i.e., positive, zero or negative) curvature and $M_3$ to be a product manifold $M_2 \times S^1$ as stated above. We assume that all $p_i=6$ sources wrap $\kappa_3$ and those with $p_i=8$ wrap in addition $M_2$.
We furthermore set $F_4=0$ but allow non-zero
\begin{equation}
F_0, \qquad F_2 \sim \text{dvol}_{M_2}, \qquad H_3\sim \text{dvol}_{M_3}. \label{fluxansatz0}
\end{equation}

We will keep track of three moduli: the dilaton modulus $\tau=\e^{-\phi}$
and the two volume moduli of $\kappa_3$ and $M_3$, which we denote by $\alpha_k$, $\alpha_m$ and define such that
\begin{equation}
g_{xy} = \alpha_k \hat g_{xy}, \qquad g_{ab} = \alpha_m \hat g_{ab}. \label{fr}
\end{equation}
Here $x,y$ are indices on $\kappa_3$ and $a,b$ are indices on $M_3$. The hatted metrics are fiducial metrics with a fixed volume (normalized, e.g., to unity).

Denoting by ${\R}_{\kappa_3}$, ${\R}_{M_3}$ the scalar curvatures of $\kappa_3$ and $M_3$, the possible 4-derivative terms compatible with our ansatz are
\begin{align}
\mathcal{L}_{\alpha^{\prime 2},i} &= c_{1i} (\e^{\phi} F_0)^4 + c_{2i}^{(1)} (\e^{\phi}F_0)^2 \mathcal{R}_{\kappa_3}
+ c_{2i}^{(2)} (\e^{\phi}F_0)^2 \mathcal{R}_{M_3} + c_{3i}^{(1)} \mathcal{R}_{\kappa_3}^2 + c_{3i}^{(2)} \mathcal{R}_{M_3}^2 \nl
+ c_{3i}^{(3)} \mathcal{R}_{\kappa_3}\mathcal{R}_{M_3}
+ c_{4i} (|{H}_3|^2)^2 + c_{5i}^{(1)} |H_3|^2 \mathcal{R}_{\kappa_3} + c_{5i}^{(2)} |H_3|^2 \mathcal{R}_{M_3} \nl + c_{6i} (\e^{\phi} F_0)^2 |H_3|^2
+ c_{7i} (\e^{2\phi} |F_2|^2)^2 + c_{8i} \e^{4\phi} F_0^2 |F_2|^2
+ c_{9i}^{(1)} \e^{2\phi} |F_2|^2 \mathcal{R}_{\kappa_3} \nl
+ c_{9i}^{(2)} \e^{2\phi} |F_2|^2 \mathcal{R}_{M_3} + c_{10i} \e^{2\phi} |F_2|^2 |H_3|^2. \label{sgsglsgji}
\end{align}
Here we used that, for 2D manifolds and 3D Einstein manifolds, any correction involving Riemann tensors can be rewritten in terms of the scalar curvature.

Performing a dimensional reduction of the type IIA action as in Section \ref{sec:scal} and using our ansatz \eqref{dfjogrdj}--\eqref{sgsglsgji}, we obtain the scalar potential
\begin{align}
	V=&\frac{1}{\tau^2 \alpha_k^{3/2} \alpha_m^{3/2}} \Bigg[-\frac{\hat{\mathcal{R}}_{\kappa_3}}{\alpha_k} -\frac{\hat{\mathcal{R}}_{M_3}}{\alpha_m} +\frac{F_0^2}{2 \tau^2} +\frac{|\hat F_2|^2}{2 \tau^2\alpha_m^2} +\frac{|\hat{H}_3|^2}{2 \alpha_m^3} + \sum_{i} \frac{T_i }{2 \pi \hat V_i \tau \alpha_m^{\frac{9-p_i}{2}}} \nn \\&
    \Bigg(1 + c_{1i} \frac{F_0^4}{\tau^{4}} + \frac{F_0^2}{\tau^2}\left[c_{2i}^{(1)} \frac{\hat{\R}_{\kappa_3}}{\alpha_k} + c_{2i}^{(2)} \frac{\hat{\R}_{M_3}}{\alpha_m}\right] + c_{3i}^{(1)}\frac{\hat{\mathcal{R}}^2_{\kappa_3}}{\alpha_k^2}+c_{3i}^{(2)}\frac{\hat{\mathcal{R}}^2_{M_3}}{\alpha_m^2}+c_{3i}^{(3)} \frac{ \hat{\mathcal{R}}_{\kappa_3}\hat{\mathcal{R}}_{M_3}}{\alpha_k \alpha_m} \nn \\&
    + c_{4i} \frac{(|\hat{H}_3|^2)^2}{\alpha_m^{6}}
    + \frac{|\hat{H}_3|^2}{\alpha_m^3}\left[c_{5i}^{(1)} \frac{\hat{\R}_{\kappa_3}}{\alpha_k} + c_{5i}^{(2)} \frac{\hat{\R}_{M_3}}{\alpha_m}\right]
    + c_{6i} \frac{F_0^2|\hat{H}_3|^2}{\tau^{2}\alpha_m^{3}} + c_{7i} \frac{(|\hat F_2|^2)^2}{\tau^4\alpha_m^4} \nn \\& + c_{8i} \frac{F_0^2 |\hat F_2|^2}{\tau^4\alpha_m^2} + \frac{|\hat F_2|^2}{\tau^2\alpha_m^2} \left[ c_{9i}^{(1)} \frac{\hat{\mathcal{R}}_{\kappa_3} }{\alpha_k}
    + c_{9i}^{(2)} \frac{\hat{\mathcal{R}}_{M_3} }{\alpha_m}\right] + c_{10i} \frac{|\hat F_2|^2 |\hat H_3|^2}{\tau^2\alpha_m^5}    
    \Bigg) \Bigg], \label{dsjlsdfljglj}
\end{align}
where $p_i=6,8$ depending on the source.

\subsubsection{Classical no-go}
\label{sec:o6class2}

In the classical case $c_{ai}^{(b)}=0$, the scalar potential simplifies to
\begin{equation}
\label{eq:PotO6O8}
V=\frac{1}{\tau^2 \alpha_k^{3/2} \alpha_m^{3/2}} \left[-\frac{\hat{\mathcal{R}}_{\kappa_3}}{\alpha_k} -\frac{\hat{\mathcal{R}}_{M_3}}{\alpha_m} +\frac{F_0^2}{2 \tau^2} +\frac{|\hat F_2|^2}{2 \tau^2\alpha_m^2} +\frac{|\hat{H}_3|^2}{2 \alpha_m^3} + \sum_{i} \frac{T_i}{2 \pi \hat V_i \tau \alpha_m^{\frac{9-p_i}{2}}} \right].
\end{equation}
Using the equation of motion $\partial_{\alpha_k}V=0$, we find on-shell:
\begin{equation}
V = \frac{2 \hat{\mathcal{R}}_{\kappa_3}}{3 \tau^2 \alpha_k^{5/2} \alpha_m^{3/2}} = - \frac{4}{15}\alpha_k^2 \del^2_{\alpha_k} V. \label{ov}
\end{equation}
This yields AdS for $\hat{\mathcal{R}}_{\kappa_3}<0$, Minkowski for $\hat{\mathcal{R}}_{\kappa_3}=0$ and unstable dS for $\hat{\mathcal{R}}_{\kappa_3}>0$, while meta-stable dS solutions are ruled out. This is true for all ``CDT2-like'' models that fit into the general ansatz described above. Note that further model-dependent constraints arise from imposing the orientifold involution, tadpole cancelation and the equations of motion for the remaining moduli, which we do not analyze here.

\subsubsection{Adding $\alpha^\prime$ corrections}
\label{sec:o6alpha2}

We now add the effect of the $\alpha'$ corrections to the O-planes/D-branes. Using \eqref{dsjlsdfljglj}, we find
\begin{align}
V + \frac{2}{3} \alpha_k \del_{\alpha_k} V = \frac{2 \hat{\R}_{\kappa_3}}{3 \tau^2 \alpha_k^{5/2} \alpha_m^{3/2}} \Bigg[& 1
    - \sum_{i} \frac{T_i }{2 \pi \hat V_i \tau \alpha_m^{\frac{9-p_i}{2}} } \left( c_{2i}^{(1)} \frac{F_0^2}{\tau^2} + 2 c_{3i}^{(1)} \frac{\hat{\R}_{\kappa_3}}{\alpha_k} \right. \notag \\ & + \left. c_{3i}^{(3)} \frac{\hat{\R}_{M_3}}{\alpha_m} + c_{5i}^{(1)}\frac{|\hat{H}_3|^2}{\alpha_m^3} + c_{9i}^{(1)} \frac{|\hat F_2|^2}{\tau^2\alpha_m^2} \right)  \Bigg] \label{sdgsdlgs}
\end{align}
and
\begin{align}
\alpha_k^2 \del^2_{\alpha_k} V + \frac{5}{2} \alpha_k \del_{\alpha_k} V = -\frac{5 \hat{\R}_{\kappa_3}}{2 \tau^2 \alpha_k^{5/2} \alpha_m^{3/2}} \Bigg[& 1
    - \sum_{i} \frac{T_i }{2 \pi \hat V_i \tau \alpha_m^{\frac{9-p_i}{2}}} \left(c_{2i}^{(1)} \frac{F_0^2}{\tau^2} + \frac{14}{5} c_{3i}^{(1)} \frac{\hat{\R}_{\kappa_3}}{\alpha_k} \right. \notag \\ & + \left. c_{3i}^{(3)} \frac{\hat{\R}_{M_3}}{\alpha_m} + c_{5i}^{(1)}\frac{|\hat{H}_3|^2}{\alpha_m^3} + c_{9i}^{(1)} \frac{|\hat F_2|^2}{\tau^2\alpha_m^2} \right) \Bigg]. \label{sdgsdlgs2}
\end{align}
Note that the corrections proportional to $c_{1i}$, $c_{4i}$, $c_{6i}$, $c_{7i}$, $c_{8i}$ and $c_{10i}$ disappear completely in the above equations. They therefore neither affect the vacuum energy nor the stability of $\alpha_k$. Using the equation of motion $\partial_{\alpha_k}V=0$, it follows from the above equations that on-shell
\begin{equation}
V = -\frac{4}{15}\alpha_k^2 \del^2_{\alpha_k} V + \frac{8}{15} \sum_{i} \frac{T_i }{2 \pi \hat V_i} \frac{c_{3i}^{(1)} \hat{\R}_{\kappa_3}^2}{\tau^3 \alpha_k^{7/2} \alpha_m^{\frac{12-p_i}{2}}}. \label{sdjgis}
\end{equation}
We thus see that, for $c_{3i}^{(1)}=0$, we get the same no-go result as in the classical case. In particular, we then have $\partial_{\alpha_k}^2 V < 0$ whenever $V>0$ and therefore no meta-stable dS is possible.

Including the $c_{3i}^{(1)}$ correction, we can naively evade the no-go if $\sum_i \frac{T_ic_{3i}^{(1)}}{\hat V_i \alpha_m^{\frac{12-p_i}{2}}}>0$.
However, the problem with this idea is that the classical vacuum energy does not vanish for $\hat{\mathcal{R}}_{\kappa_3}\neq 0$ according to \eqref{ov}. Therefore, in order to evade the no-go, some of the correction terms in \eqref{sdgsdlgs} and \eqref{sdgsdlgs2}
have to be of the same order as the classical vacuum energy. For concreteness, let us assume that this is the case for the $c_{3i}^{(1)}$ term (the discussion for the other corrections is analogous):
\begin{equation}
\frac{\hat{\mathcal{R}}_{\kappa_3}}{\tau^2 \alpha_k^{5/2} \alpha_m^{3/2}} \sim \sum_{i} \frac{T_i }{2 \pi \hat V_i} \frac{c_{3i}^{(1)} \hat{\R}_{\kappa_3}^2}{\tau^3 \alpha_k^{7/2} \alpha_m^{\frac{12-p_i}{2}}}.
\end{equation}
Going back to the 10D string-frame metric (the one without the hats as defined in \eqref{fr}), this corresponds to
\begin{equation}
\R_{\kappa_3} \sim \sum_{i} \e^\phi \frac{T_i }{2 \pi V_i} c_{3i}^{(1)} \R_{\kappa_3}^2. \label{fslgkfs}
\end{equation}

This is problematic since the 4-derivative correction can only compete with the classical 2-derivative term at large curvature.
We thus expect that an infinite number of further higher-derivative corrections becomes comparable to the ``leading'' ones and the $\alpha^\prime$ expansion breaks down.
We stress that the right-hand side of \eqref{fslgkfs} has support on the \emph{smeared} source term (or, equivalently, the zero mode of the localized one). Avoiding the dS no-go would therefore require large curvature everywhere on the 10D spacetime, not just very close to the O-planes (where large curvature is expected and not necessarily problematic, as explained in Section \ref{sec:smear}).
To be more specific, one can verify using the equations of motion that $\sum_i\e^\phi\frac{T_i }{2 \pi V_i}$ is at most of the order of the classical 2-derivative terms, i.e., of $\R_{\kappa_3}$, $\R_{M_3}$, $|H_3|^2$, $\e^{2\phi}|F_q|^2$.
Furthermore, $c_{3i}^{(1)} = \frac{\mathcal{O}(1)}{48\cdot 16\pi^2} = \mathcal{O}(10^{-4})$ for both D-branes and O-planes up to the usual field-redefinition ambiguities \cite{Bachas:1999um, Robbins:2014ara}. Estimating $\sum_{i} \frac{T_i }{V_i} c_{3i}^{(1)} = \mathcal{O}(10^{-4})\sum_{i} \frac{T_i }{ V_i}$ in \eqref{fslgkfs}, we conclude that avoiding the no-go would require $\R_{\kappa_3}$, $\R_{M_3}$, $|H_3|^2$ and/or $\e^{2\phi}|F_q|^2$ to be of the order $\frac{1}{\sqrt{c_{3i}^{(1)}}} \sim \mathcal{O}(10^2)$ in string units.

In conclusion, taking into account 4-derivative corrections to the source terms does not allow to avoid the dS no-go for the CDT2-like models in the regime where supergravity can be trusted. A caveat of our results in this section is that we did not include possible backreaction effects. As discussed before, the motivation to consider an approximately smeared regime is that it guarantees that there are no large singular holes and supergravity is reliable. However, a smeared regime may not always be required, especially in compactifications with O8-planes where the backreaction can in some cases be strong without creating any holes (see Fig.~\ref{smeared2}). It would therefore be interesting to study such backreaction effects further in the CDT2-like models.
However, note that, whenever we did include backreaction effects in this paper, we found that they do not affect the analysis.
In particular, this was true for the original CDT2 model, where our no-go for the fully backreacted ansatz was in full agreement with the corresponding smeared argument.

\section{Conclusions}
\label{sec:concl}

In this paper, we studied the possibility of dS vacua in flux compactifications with O8/D8 and O6/D6 sources, in particular in the CDT1 and CDT2 models proposed in \cite{Cordova:2018dbb, Cordova:2019cvf} and certain variants thereof with a similar source and flux content.

\begin{table}[t]\renewcommand{\arraystretch}{1.2}\setlength{\tabcolsep}{6pt}
\begin{tabular}{|l|l|l|}
\hline 
model & classical & classical + 4-derivative terms \\
\hline 
CDT1 & smeared, backreacted (Sec.~\ref{sec:o8nogo} and \cite{Cribiori:2019clo}) & smeared (Sec.~\ref{sec:o8nogo2}) \\
CDT1-like & smeared, backreacted (Sec.~\ref{sec:o8nogo} and \cite{Cribiori:2019clo}) & smeared (Sec.~\ref{sec:o8nogo2}) \\
CDT2 & smeared (Sec.~\ref{sec:o6smear}), & smeared (Sec.~\ref{sec:o6smear}), \\
 & backreacted (Sec.~\ref{sec:o6class}) & backreacted (Sec.~\ref{sec:o6alpha}) \\
CDT2-like & smeared (Sec.~\ref{sec:o6class2}) & smeared (Sec.~\ref{sec:o6alpha2}) \\
\hline
\end{tabular}
\\

\caption{Summary of our no-go arguments against dS in the CDT1 and CDT2 models and variants thereof with links to the sections where they are discussed. By ``CDT1-like'' we mean models with O8/D8 sources and $F_0$ flux that fit into our ansatz in Section \ref{sec:o8}. By ``CDT2-like'', we mean models with O6/D6/O8/D8 sources and $F_0$, $F_2$ and $H_3$ fluxes that fit into our ansatz of Section \ref{sec:o6gen}.
}

\label{tab1}
\end{table}

We found that none of these models admit \emph{classical} dS vacua, i.e., dS vacua which arise at the level of the classical equations of motion with classical source terms (as stated in App.~\ref{app:eom0}). For the CDT1 model and CDT1-like variants, we reviewed a no-go theorem of \cite{Cribiori:2019clo} showing this. We furthermore derived new no-gos for the CDT2 model and CDT2-like variants which imply that at best AdS/Minkowski vacua or unstable dS solutions are allowed classically in these models.

On the other hand, the numerical results of \cite{Cordova:2018dbb, Cordova:2019cvf} suggest that dS solutions become possible in the CDT1 and CDT2 models if one relaxes the assumption of classical source terms. Our goal was to do this in such a way that we retain control over the $\alpha^\prime$ expansion and keep the simplicity of the original models. This motivated us to consider an ``almost classical'' scenario in which the leading $\alpha^\prime$ corrections to the classical source terms at the 4-derivative level are taken into account but higher-order corrections to the sources as well as corrections in the bulk are still self-consistently neglected. We worked out how the 4-derivative corrections modify the classical no-go results and studied whether this is sufficient to obtain dS vacua. Unfortunately, we found that dS vacua are still not possible including the 4-derivative terms, neither in the original CDT1 and CDT2 models nor in the generalized models we considered.
Our various no-go results for the different models are summarized in Table \ref{tab1}.

As explained before, an important assumption in this paper (and in the earlier work \cite{Cribiori:2019clo}) is that we are in a regime where the $\alpha^\prime$ expansion is reliable on most of the 10D spacetime. In particular, we imposed that the O-planes are ``thin'' in the sense that any singular holes in their vicinity in which supergravity breaks down must be small compared to the size of the compact space (Small-Hole Condition). In such a regime, we expect that the holes do not have a significant effect on low-energy properties like the vacuum energy so that we can reliably compute it using classical supergravity with classical source terms (plus possibly the next-to-leading $\alpha^\prime$ corrections in case the leading result vanishes) whereas non-perturbative string theory is only required for the small-distance physics very close to the O-planes.
In particular, some of our arguments in this paper and in \cite{Cribiori:2019clo} integrate the classical equations of motion over the internal space to constrain the vacuum energy, which was criticized in \cite{Cordova:2019cvf} to be unreliable in the presence of holes. However, we argued using limits of certain boundary integrals that our approach is consistent in the small-hole regime and that the effect of the hole regions in this regime is correctly captured by the  classical tensions of the sources and their leading $\alpha^\prime$ corrections. We therefore believe that the criticism of \cite{Cordova:2019cvf} is unjustified in the small-hole regime.

On the other hand, in a regime where the O-planes become ``thick'', i.e., the holes created by their backreaction are so large that supergravity breaks down on a large part of the spacetime, we should not expect that supergravity can be used to compute the vacuum energy. In such a large-hole regime, the calculations in this paper and in \cite{Cordova:2018dbb, Cribiori:2019clo, Cordova:2019cvf} are not reliable and one should instead use truly non-perturbative methods to study string vacua.

A further assumption used in some (but not all) of our no-go arguments is the existence of an approximately smeared regime. As discussed in Section \ref{sec:smear}, such a regime often arises as an effective behavior of a solution for small enough $g_s$ where backreaction effects are negligible on most of the spacetime. This is attractive since it ensures that the holes surrounding the O-planes are small while at the same time making the solutions very simple.
However, there may also be situations where a solution is not approximately smeared but still reliably described by supergravity, in particular if it contains O$8^-$-planes or positive-tension objects, which can backreact strongly without creating large holes.
In this paper, some of our no-go arguments were derived including the full backreaction of all sources, but we relied on a smeared regime in those no-gos which are based on identifying a tachyon, see Table \ref{tab1}. It would be very interesting to derive the backreacted analogues of the latter no-gos in future work, which would require computing the corresponding warped effective field theory for each model.
Let us note here that, in those cases where we were able to take into account the full backreaction, we found that the no-gos are \emph{not} affected by backreaction effects but in full agreement with the corresponding smeared results.
More results taking into account the backreaction in the CDT1 model will be provided in a separate paper \cite{Junghans:2024}.

Although we ruled out the almost-classical-dS scenario under the stated assumptions in the models we considered, it would be interesting to study the scenario further in other flux compactifications of type IIA/B string theory.
In order that 4-derivative corrections to the O-plane/D-brane tensions can have a leading effect in the perturbative regime where curvatures/energy densities are small on most of the spacetime, the scenario requires classically either a Minkowski solution or one where several classical terms in the vacuum energy almost cancel out. Furthermore, the 4-derivative terms must have the right form to lift the vacuum energy to a positive value without destabilizing the solution. We leave a general analysis of this scenario and possible no-gos for future work.

We finally note that the obstructions to dS found in this paper are conceptually very similar to those arising in various other dS scenarios in string theory.
In particular, various recent results suggest that dS vacua may in general not be allowed in perturbative regimes of string theory where the scalar potential is self-consistently approximated by a few leading terms in the $\alpha^\prime$ expansion \cite{Junghans:2018gdb, Banlaki:2018ayh, Carta:2019rhx, Gao:2020xqh, Blumenhagen:2022dbo, Junghans:2022exo, Gao:2022fdi, Junghans:2022kxg, Hebecker:2022zme, Schreyer:2022len, Schreyer:2024pml, ValeixoBento:2023nbv, Junghans:2023lpo}. It will be important to gain a better understanding of this observation and its possible consequences for the nature of dark energy.

\section*{Acknowledgments}

This research was funded by the Austrian Science Fund (FWF) under project number P 34562-N and by the German Research Foundation (Deutsche Forschungsgemeinschaft) under project number 516370439.

\begin{appendix}

\section{Equations of motion}
\label{app:eom0}

In this appendix, we state the equations of motion that follow from the variation of \eqref{action}, \eqref{action2} at the classical level (i.e., for $\mathcal{L}_{\alpha^{\prime 2},i}=0$). The dilaton and Einstein equations are
\begin{align}
0 &= -8 \e^{\phi} \nabla^2 \e^{-\phi} -4d\e^{-A}\nabla^2 \e^A  -2d(d-1)( \partial A)^2 + 8d ( \partial A \cdot \partial \phi) + 2\e^{-2A} \mathcal{ R}_d \nl + 2\mathcal{ R}_{10-d} - |H_3|^2 - \sum_i \frac{T_{i}}{2\pi} \e^\phi \delta(\Sigma_i), \label{eom1} \\
0 &=  \frac{2d}{8}\e^{\phi}\nabla^2\e^{-\phi} +d\e^{-A}\nabla^2\e^A +\frac{2d}{8}( \partial \phi)^2+d(d-1)( \partial A)^2 -\frac{2d(8+d)}{8}( \partial A \cdot \partial \phi) \nl - \e^{-2A} \mathcal{R}_d - \frac{d}{8} |H_3|^2 - \sum_{q} \frac{(q-1)d}{16}\e^{2\phi} |F_q|^2 - \sum_i \frac{T_{i}}{2\pi} \frac{(7-p_i)d}{16} \e^\phi \delta(\Sigma_i), \label{eom2} \\
0 &= 2\e^{\phi} \nabla_m\partial_n \e^{-\phi} + d\e^{-A}\nabla_m\partial_n \e^A + \frac{1}{4} g_{mn}\e^{\phi} \nabla^2\e^{-\phi} + \frac{1}{4} g_{mn} (\partial\phi)^2  -\frac{d}{4} g_{mn}(\partial A\cdot\partial\phi) \nl - 2(\partial_m\phi)(\partial_n\phi)  -\mathcal{R}_{mn}
+ \frac{1}{2}|H_3|^2_{mn} - \frac{1}{8}g_{mn}|H_3|^2 + \frac{1}{2} \e^{2\phi} \sum_{q} \left( |F_q|^2_{mn}- \frac{q-1}{8}g_{mn}|F_q|^2 \right) \nl - \frac{1}{2}\sum_i \frac{T_{i}}{2\pi} \left(\Pi^{(i)}_{mn}-\frac{p_i+1}{8}g_{mn}\right) \e^\phi \delta(\Sigma_i), \label{eom3}
\end{align}
where $\Pi^{(i)}_{mn}=g_{mn}$ for indices parallel to the corresponding source and $\Pi^{(i)}_{mn}=0$ for transverse indices. Note that $q$ runs over the even numbers between $0$ and $10-d$ and we dualized all spacetime-filling terms in the RR field strengths into internal ones using the duality constraint stated below \eqref{action}.
The covariant derivative $\nabla_m$ and the Laplacian $\nabla^2$ are defined with respect to the $(10-d)$-dimensional metric $g_{mn}$ and the delta distributions are defined as
\begin{equation}
\int \d^{10} x \sqrt{-g_{10}}\, f \delta(\Sigma_i) = \int_{\Sigma_i} \d^{p_i+1} x \sqrt{-g_{p_i+1}}\, f|_{\Sigma_i}, \label{delta}
\end{equation}
where $f$ is a function of the internal coordinates and $f|_{\Sigma_i}$ is its restriction to $\Sigma_i$. This implies in particular that $\delta(\Sigma_i) = \frac{\delta(\vec x)}{\sqrt{g_{9-p_i}}}$ in local coordinates, including an inverse metric determinant.

The equations of motion and Bianchi identities for the RR/NSNS form fields are
\begin{align}
&\d \left(\star_{10} F_q\right) = - H_3 \w \star_{10} F_{q+2}, \qquad \d F_q = H_3 \w F_{q-2} - (-1)^{q(q+1)/2} \sum_{i}\frac{Q_{i}}{2\pi} \delta_{9-p_i}, \label{eom4} \\
&\d \left(\e^{-2\phi}\star_{10} H_3\right) = -\sum_q \star_{10} F_q \w F_{q-2}, \qquad \d H_3 =0, \label{eom5}
\end{align}
where $\delta_{9-p_i} = \delta(\Sigma_i) \text{dvol}_{9-p_i}$ and $Q_{i}=\pm T_{i}$ (the upper sign is for branes/O-planes and the lower one for anti-branes/anti-O-planes). The sum in \eqref{eom4} runs over all $i$ with $p_i=8-q$.

\section{O6$^-$ boundary conditions in the CDT2 model}
\label{app:o6}

In this appendix, we determine the boundary conditions near the O$6^-$-planes in the dS solution of \cite{Cordova:2019cvf}.

\subsection{Ansatz and equations of motion}
\label{app:eom}

We consider the ansatz stated in Section \ref{sec:o6ansatz}, which we repeat here for convenience. The metric in the string frame is
\begin{equation}
    \d s_{10}^2= \e^{2A}\d s^2_{4}+\e^{-2A}\left(\e^{2\lambda_3}\d s_{\kappa_3}^2+\e^{2\lambda_2}\d s_{S^2}^2 + R^2\d z^2\right),
\end{equation}
where the warp factor $A$, along with the dilaton $\phi$ and $\lambda_i$, are functions of the $z$ coordinate and $\d s_{\kappa_3}^2$ is the metric of a negatively curved Einstein manifold with Ricci scalar $\mathcal{R}_{\kappa_3}$. The (anti-)O$6^-$-planes sit at $z=0$ and $z=\pi$ and the O$8^+$-plane sits at $z=\frac{\pi}{2}$ in our conventions, where the orientifold involution implies that the solution at $z-\frac{\pi}{2}<0$ is a copy of the one at $z-\frac{\pi}{2}>0$. We can therefore focus on the O6 at $z=0$.
We further set
\begin{equation}
    F_2=f_2 \text{dvol}_{S^2},\qquad H_3=\frac{f_2'}{F_0} \d z\wedge\text{dvol}_{S^2},\qquad F_4=0 \label{slsfgjl}
\end{equation}
with $'=\frac{\d}{\d z}$ and non-vanishing $F_0$. This ansatz follows from substituting \eqref{fluxansatz} into the Bianchi identity $\d F_2 =F_0H_3$ at points without localized sources. One verifies that \eqref{slsfgjl} solves the remaining Bianchi identities and the equations of motion for the fluxes if
\begin{equation}\label{eom:f2}
    f_2''=\e^{-2A+2\phi}f_2 F_0^2 R^2 +f_2'\left(-4A'+2\lambda_2'-3\lambda_3'+2\phi'\right)
\end{equation}
holds away from the sources.

Let us also state the Einstein and dilaton equations:
\begin{align}
        2R^2 \mathcal{R}_4 \e^{-4A} ={}& \e^{4A-4\lambda_2}\left(-\frac{(f_2')^2}{F_0^2}+f_2^2 R^2 \e^{-2A+2\phi}+F_0^2 R^2 \e^{-6A+4\lambda_2+2\phi}\right)+8(\phi')^2-16\lambda_2'\phi' \notag \\
        & -24\lambda_3'\phi'+4(\lambda_2')^2+12(\lambda_3')^2+24\lambda_2'\lambda_3'+8A'\phi'-16(A')^2 \notag \\
        & -2R^2\mathcal{R}_{\kappa_3}\e^{-2\lambda_3}-4R^2\e^{-2\lambda_2}, \label{eom:cc}\\
        16A'' ={}& \e^{4A-4\lambda_2}\left(-\frac{2(f_2')^2}{F_0^2}+6f_2^2 R^2 \e^{-2A+2\phi}+6F_0^2 R^2 \e^{-6A+4\lambda_2+2\phi}\right)+16(\phi')^2 \notag \\   &-32\lambda_2'\phi'-48\lambda_3'\phi'+8(\lambda_2')^2+24(\lambda_3')^2+48\lambda_2'\lambda_3'-32\lambda_2'A'-48\lambda_3'A' \notag \\
        & +48A'\phi'-32(A')^2-4R^2\mathcal{R}_{\kappa_3}\e^{-2\lambda_3}-8R^2\e^{-2\lambda_2} , \label{eom:W}\\
        8\lambda_2'' ={}& \e^{4A-4\lambda_2}\left(-\frac{5(f_2')^2}{F_0^2}+f_2^2R^2 \e^{-2A+2\phi}+5F_0^2R^2 \e^{-6A+4\lambda_2+2\phi}\right)+8(\phi')^2 -24\lambda_3'\phi' \notag \\
        & -12(\lambda_2')^2+12(\lambda_3')^2+8A'\phi'-16(A')^2-2R^2\mathcal{R}_{\kappa_3}\e^{-2\lambda_3}+4R^2\e^{-2\lambda_2} , \label{eom:lambda2}\\
        8\lambda_3'' ={}& \e^{4A-4\lambda_2}\left(-\frac{(f_2')^2}{F_0^2}+5f_2^2R^2 \e^{-2A+2\phi}+5F_0^2R^2 \e^{-6A+4\lambda_2+2\phi}\right)+8(\phi')^2 \notag \\
        & -16\lambda_2'\phi'-8\lambda_3'\phi'+4(\lambda_2')^2-12(\lambda_3')^2+8\lambda_2'\lambda_3'+8A'\phi'-16(A')^2 \notag \\
        & +\frac{2}{3}R^2\mathcal{R}_{\kappa_3}\e^{-2\lambda_3}-4R^2\e^{-2\lambda_2} , \label{eom:lambda3}\\
        4\phi'' ={}& \e^{4A-4\lambda_2}\left(-\frac{2(f_2')^2}{F_0^2}+3f_2^2R^2 \e^{-2A+2\phi}+5F_0^2R^2 \e^{-6A+4\lambda_2+2\phi}\right)+8(\phi')^2  -8\lambda_2'\phi' \notag \\ & -12\lambda_3'\phi', \label{eom:dil}
\end{align}
where $\mathcal{R}_4$ is the external curvature and we limit ourselves to regions away from sources.

\subsection{Leading-order behavior}
\label{sec:lo_behaviour}

Following a similar computation in \cite[App.~B]{Blaback:2011pn}, we now compute the behavior of the 10D supergravity fields near the O6 at $z=0$. Our ansatz at leading order in $z$ is
\begin{gather}
\e^A = k_0 z^K+\ldots, \qquad \e^\phi = f_0 z^F+\ldots, \qquad \e^{\lambda_2} = m_0 z^{M}+\ldots, \notag \\ \e^{\lambda_3} = n_0 z^{N}+\ldots, \qquad f_2 = l_0 z^{L}+\ldots, \label{eq:O6fields1}
\end{gather}
where we assume $k_0, f_0, m_0, n_0, l_0 \in \mathbb{C}$ and $K, F, M, N, L \in \mathbb{R}$.
According to footnote 8 in \cite{Cordova:2019cvf}, the warp factor $\e^{-4A}$ does not diverge and $\e^\phi$, $\e^{\lambda_2}$, $\e^{\lambda_3}$ tend to zero at $z=0$. We therefore impose
\begin{equation}
K \le 0, \qquad F, M,N>0. \label{exponents}
\end{equation}
Since the warp factor $\e^{-4A}$ and the other functions can be negative inside the hole for orientifolds, the coefficients $k_0, f_0, m_0, n_0, l_0$ do not have to be positive or even real. In the above expansion, we ignore terms of higher order but assume that those shown there are the actual leading order, that is $k_0, f_0, m_0, n_0, l_0$ do not vanish.

Derivatives of the warp factor satisfy
\begin{equation}\label{eq:O6fields2}
    \begin{gathered}
        A'=Kz^{-1}+\ldots,\qquad A''=-K z^{-2}+\ldots,
    \end{gathered}
\end{equation}
and similar expressions hold for derivatives of $\phi$, $\lambda_2$ and $\lambda_3$.

Inserting \eqref{eq:O6fields1} into \eqref{eom:f2}, we obtain the relation
\begin{equation}\label{eq:f2expansion}
    L\left(2F-4K-L+2M-3N +1\right)=0 ,
\end{equation}
which will be used below. In the following, we derive further constraints on the exponents in \eqref{eq:O6fields1}, where we distinguish three separate cases.

\subsubsection*{Case 1: $L>2M-2K>0$}

At leading order (LO) in $z$, \eqref{eom:dil} yields
\begin{equation}
4F(1 + 2F - 2M - 3N) = 0.
\end{equation}
Using this together with \eqref{eq:f2expansion}, we obtain
\begin{equation}
L = 4M-4K, \qquad N = \frac{1}{3}(2F-2M+1).
\end{equation}
The possible LO terms of \eqref{eom:cc}--\eqref{eom:lambda3} are then
\begin{align}
0 &= -\frac{4 R^2}{m_0^2} \begin{pmatrix} 1 \\ 1 \\ -1 \\ 1 \end{pmatrix}
 z^{-2M}-\frac{2R^2\mathcal{R}_{\kappa_3}}{n_0^2} \begin{pmatrix} 1 \\ 1 \\ 1 \\ -\frac{1}{3} \end{pmatrix} z^{-\frac{2}{3}(2F-2M+1)} \nl + \left(-\frac{8}{3}F^2+8KF+\frac{16}{3}FM-\frac{8}{3}F-16K^2-\frac{20}{3}M^2+\frac{8}{3}M+\frac{4}{3} \right) z^{-2}.
\end{align}
This can only be solved at LO if
\begin{equation}
M<1, \quad F-M <1, \quad -\frac{8}{3}F^2+8KF+\frac{16}{3}FM-\frac{8}{3}F-16K^2-\frac{20}{3}M^2+\frac{8}{3}M+\frac{4}{3} = 0. \label{fskgfkgjf}
\end{equation}
Solving \eqref{fskgfkgjf} for $F>0$, $K\le 0$ yields
\begin{equation}
K = \frac{F}{4} - \frac{\sqrt{-15F^2+48FM-60M^2-24F+24M+12}}{12}.
\end{equation}
This is non-positive for $M-\frac{1}{2}-\frac{\sqrt{3-6M^2}}{2} \le F \le M-\frac{1}{2}+\frac{\sqrt{3-6M^2}}{2}$, $M\le \frac{1}{\sqrt{2}}$. Because of $N>0$, we further have to satisfy $F>M-\frac{1}{2}$. Putting everything together, we find that the allowed boundary conditions are
\begin{align}
L &= 4M -F + \frac{\sqrt{-15F^2+48FM-60M^2-24F+24M+12}}{3}, \\ N &= \frac{1}{3}(2F-2M+1), \\ K &= \frac{F}{4} - \frac{\sqrt{-15F^2+48FM-60M^2-24F+24M+12}}{12}
\end{align}
in the intervals
\begin{equation}
\text{max}\left(0,M-\frac{1}{2}\right) < F \le M-\frac{1}{2}+\frac{\sqrt{3-6M^2}}{2}, \qquad 0< M < \frac{1}{\sqrt{2}}.
\end{equation}
Note that further constraints on the exponents might arise from the equations of motion at subleading orders in the $z$ expansion or due to global constraints which we do not study here.

\subsubsection*{Case 2: $L=2M-2K>0$}

Using \eqref{eq:f2expansion}, we obtain
\begin{equation}
N = \frac{1}{3}(2F-2K+1).
\end{equation}
Taking \eqref{eom:cc} minus $\frac{1}{2}\cdot$\eqref{eom:W} then yields
\begin{equation}
-16K^2 + 16KM=0.
\end{equation}
The only solution consistent with $K\le 0$, $M>0$ is
\begin{equation}
K=0.
\end{equation}
The possible LO terms of \eqref{eom:lambda3} minus $\frac{1}{2}\cdot$\eqref{eom:W} are
\begin{equation}
\frac{8 R^2\mathcal{R}_{\kappa_3}}{3n_0^2} z^{-\frac{2}{3}(2F+1)} - \frac{16}{3}(2F+1)M z^{-2}=0.
\end{equation}
Since $F>0$, $M>0$, this can only be solved at LO for $F=1$. The possible LO terms of \eqref{eom:lambda2} minus $\frac{1}{2}\cdot$\eqref{eom:W} minus $2\cdot$\eqref{eom:dil} are then
\begin{equation}
\frac{8 R^2}{m_0^2} z^{-2M} - 16M(M-1) z^{-2}=0.
\end{equation}
This is not solved at LO for any $M>0$. Hence, Case 2 does not yield any consistent boundary conditions.

\subsubsection*{Case 3: $L<2M-2K$}

At LO in $z$, \eqref{eom:dil} yields
\begin{equation}
L = 0.
\end{equation}
We now consider the following combinations of the equations of motion:
\eqref{eom:cc}$-$\eqref{eom:W}$+$\eqref{eom:lambda3},
$-4\cdot$\eqref{eom:cc}$+\frac{3}{2}\cdot$\eqref{eom:W}$+$\eqref{eom:lambda2}$-2\cdot$\eqref{eom:dil},
\eqref{eom:cc}$-\frac{1}{2}\cdot$\eqref{eom:W}. The possible LO terms of these equations are
\begin{align}
0 &= \frac{8R^2\mathcal{R}_{\kappa_3}}{3n_0^2} z^{-2N} - 8 \left(2F-2M-3N+1 \right)\left(2K-N\right) z^{-2}, \label{lo1} \\
0 &= \frac{8 R^2}{m_0^2} z^{-2M} - 8 \left(2F-2M-3N+1 \right)\left(F-M-3K\right) z^{-2}, \label{lo2} \\
0 &= -\frac{2 R^2 f_0^2k_0^2l_0^2}{m_0^4} z^{2F+2K-4M} - 8 \left(2F-2M-3N+1 \right) K z^{-2}. \label{lo3}
\end{align}
The first equation can be solved at LO for $N=1 \neq \frac{1}{3}(2F-2M+1)$ or $N = \frac{1}{3}(2F-2M+1) < 1$. In the second case, all three equations are solved at LO if
\begin{equation}
N = \frac{1}{3}(2F-2M+1), \qquad M<1, \qquad 2M-1-K < F< M+1. \label{sdjksldgsdhku}
\end{equation}
The first case $N=1 \neq \frac{1}{3}(2F-2M+1)$ naively yields further solutions. Depending on whether the $z^{-2M}$ and $z^{2F+2K-4M}$ terms in \eqref{lo2} and \eqref{lo3} are assumed to be leading or subleading, we find four more possibilities to solve \eqref{lo1}--\eqref{lo3} at LO:
\begin{align}
& N=1, && F=M, && K=0, && M<1, \label{lo5} \\
& N=1, && F=\frac{7M-3}{4}, && K=\frac{M-1}{4}, && \frac{3}{7}<M<1, \\
& N=1, && M=1, && K=0, && 2\neq F>1, \\
& N=1, && M=1, && K=1-F, && 2\neq F>1. \label{lo8}
\end{align}
A further constraint is obtained from the combination $-2\cdot$\eqref{eom:cc}$+\frac{1}{2}\cdot$\eqref{eom:W}$+2\cdot$\eqref{eom:lambda2}$+3\cdot$\eqref{eom:lambda3} $-6\cdot$\eqref{eom:dil}. At LO, this yields
\begin{align}
0 &= 2F^2 + (-6K - 2M - 3N + 3)F + 8K^2 + (2M + 3N - 1)K + 2M^2 \nl + 3N^2 - 2M - 3N. \label{lo4}
\end{align}
One verifies that none of the possibilities \eqref{lo5}--\eqref{lo8} are consistent with \eqref{lo4} so that we are left with \eqref{sdjksldgsdhku} as the only possible solution. Substituting this into \eqref{lo4} yields the boundary condition
\begin{align}
L &= 0, \qquad N = \frac{1}{3}(2F-2M+1), \notag \\ K &= \frac{F}{4} - \frac{\sqrt{-15F^2+48FM-60M^2-24F+24M+12}}{12}.
\end{align}
The discussion of the allowed ranges of $F$ and $M$ proceeds as in Case 1, except that we now in addition have to satisfy $2M-1-K < F$. One verifies that this implies
\begin{align}
& \text{max}\left(0,\frac{4}{5}(2M-1) + \frac{\sqrt{-9M^2 - 6M + 9}}{10}\right) < F \le M-\frac{1}{2}+\frac{\sqrt{3-6M^2}}{2}, \\ & 0< M < \frac{1+\sqrt{6}}{5}.
\end{align}

\end{appendix}

\bibliographystyle{jhep}
\bibliography{ref}

\end{document}